\documentclass[aps,prl,twocolumn,superscriptaddress,longbibliography,reprint,floatfix]{revtex4-2}
\usepackage[T1]{fontenc}
\usepackage{amsmath,amssymb,amsfonts,amsthm,mathtools,bm}
\usepackage{graphicx}
\usepackage{booktabs}
\usepackage{placeins}
\usepackage[dvipsnames]{xcolor}
\usepackage{times}
\usepackage{enumitem}
\definecolor{myrefcolor}{rgb}{0.067,0.5,0.5}
\usepackage[breaklinks=true,colorlinks=true,linkcolor=myrefcolor,citecolor=myrefcolor,urlcolor=myrefcolor]{hyperref}
\hypersetup{pdftitle={Readout errors limit how far adaptive quantum circuits can entangle},pdfauthor={Chenfeng Cao and Ruge Lin}}
\newcommand{\ket}[1]{\lvert #1\rangle}
\newcommand{\bra}[1]{\langle #1\rvert}
\newcommand{\tr}{\operatorname{Tr}}
\newcommand{\id}{\mathcal I}
\newcommand{\Id}{\mathbb I}
\newcommand{\Prb}{\mathbb P}

\newcommand{\eps}{\varepsilon}
\newcommand{\Fe}{F_{\mathrm e}}
\newcommand{\Dopt}{D_{\eps,\mathrm{tr}}^{\star}}
\newcommand{\Topt}{T_{\eps,\mathrm{tr}}^{\star}}
\newcommand{\EB}{\mathrm{EB}}
\newcommand{\channel}{\mathcal N}
\newtheorem{theorem}{Theorem}
\newtheorem{lemma}{Lemma}
\newtheorem{corollary}{Corollary}
\begin{document}
\title{Readout errors limit how far adaptive quantum circuits can entangle}
\author{Chenfeng Cao}
\email{chenfeng.cao@fu-berlin.de}
\affiliation{Dahlem Center for Complex Quantum Systems, Freie Universit\"{a}t Berlin, 14195 Berlin, Germany}
\affiliation{HK Institute of Quantum Science \& Technology, The University of Hong Kong, Hong Kong, China}

\author{Ruge Lin}
\affiliation{The Hong Kong University of Science and Technology (Guangzhou), Guangzhou, China}

\begin{abstract}
Measurement and feedforward let quantum circuits entangle arbitrarily distant qubits in constant depth if measurement records are perfect and classical communication is instantaneous. We allow each reported bit to flip independently with a fixed probability between zero and one half, keeping all gates ideal. Consider a square-lattice device with a unique path between the endpoints and a bounded number of qubits per site. Beyond $\exp[O(D^3)]$ steps along that path, every depth-$D$ circuit starting from a product state leaves the endpoints exactly separable in every record, however rare. By copying outcomes onto spare qubits beside the path before readout, we attain a reach of $\exp[\Theta(D^3)]$ for a nondestructive parity measurement at fixed accuracy without postselection. On a one-dimensional chain, the optimal reach for the same measurement is $\exp[\Theta(D^2)]$. On any interaction graph, a constant-depth circuit starting from a product state needs $\Omega(L\ln L)$ qubits to entangle endpoints at graph distance $L$. The reliability of local readout thus limits the depth advantage of adaptive circuits for long-range entanglement.
\end{abstract}
\maketitle

Under local unitary dynamics, correlations spread within a light cone: nearest-neighbor gates acting on a chain prepared in a product state need a depth proportional to its length before the two ends can become entangled~\cite{Lieb1972,BHV2006}. Mid-circuit measurement and feedforward remove this constraint. A Bell measurement joins two short Bell pairs into a longer one, and since the joins along a chain act on disjoint qubits, all of them fit in a single parallel round~\cite{Bennett1993,Zukowski1993,Hermans2022,Baumer2024}. Each join adds two reported bits to the measurement \emph{record}, one for each of the two Pauli corrections applied at the far end of the chain. If every bit is reported correctly and classical communication is instantaneous, the far end needs only the parities of the two bit strings, and the ends become entangled in constant depth at any distance~\cite{Piroli2021,Lu2022,Tantivasadakarn2024,CaoEisert2026}. This shortcut has been used to implement long-range CNOT gates on superconducting processors~\cite{Baumer2024,Smith2026} and to prepare a toric-code ground state~\cite{FossFeig2023}.

In practice, readout is imperfect. Suppose that each reported bit is flipped independently with probability $p$, while all gates remain ideal. A wrong phase report turns $\Phi^+$ into $\Phi^-$, and a second wrong phase report undoes the first, so each correction at the far end is right only if an even number of the reports entering it are wrong. As joins accumulate, each correction becomes a coin flip. At $p=1\%$, joining 45 Bell pairs with a single readout per outcome leaves both corrections right with probability only $0.498$. Because all outcomes are equally likely a priori, no record is more trustworthy than another, and every record leaves the corrected ends in the same mixture of Bell states. A mixture of Bell states is separable exactly when no weight exceeds $1/2$, so this one is separable. Corrections based on the true outcomes would give a perfect Bell pair in every run.

Repeated readout~\cite{Hume2007} and copying onto spare qubits~\cite{Hicks2022,Kim2025,Corlett2025} supply additional reports of the same outcome, much as repeated syndrome measurements guide decoding in error correction~\cite{Dennis2002,Harper2026}. Both majority voting and selection for agreement have been tested in feedforward circuits~\cite{Koh2026}, and Smith \emph{et al.}\ raised the fidelity of a Bell pair spanning 100 intermediate qubits from $0.39(1)$ to $0.59(2)$ by postselecting on checks sensitive to readout and gate faults~\cite{Smith2026}. Using their qubit-resolved calibrations, we estimate that independent readout errors alone would lower the Bell fidelity of the unprotected circuit to about $0.52$, close to the certification threshold of $1/2$~\cite{SM}. In our Bell chain, three reports per outcome decoded by majority vote keep the averaged endpoint state entangled across almost 1500 Bell pairs instead of fewer than 45, and keeping only runs in which every triple agrees extends this beyond $400\,000$. Since each agreeing report multiplies the odds in favor of the reported value by $(1-p)/p$, a record with enough agreeing reports can be made as convincing as desired. At fixed depth, however, a circuit can gather only a limited number of reports near any one join. Could a sufficiently rare record still leave the endpoints entangled at any distance?

\begin{figure}[t]
\centering
\includegraphics[width=\columnwidth]{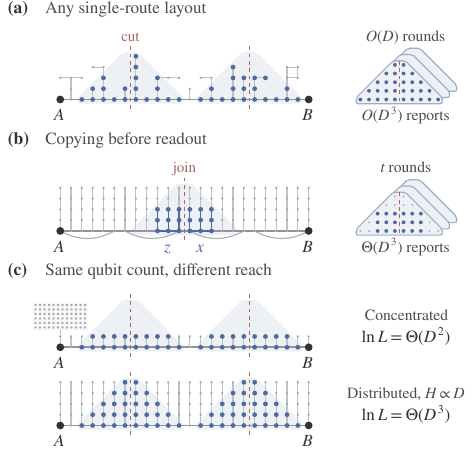}
\caption{\textbf{How geometry limits entangling reach on a single route.}
(a)~A square-lattice device with a single route (dark) of length $L$ between $A$ and $B$. Shading outlines the depth-$D$ light cones around two edges of the route (dashed cuts). Blue dots mark the neighborhood of each edge, the sites within $D$ steps of it. With a bounded number of qubits per site, the $O(D^2)$ sites of a neighborhood supply at most $O(D^3)$ reports over $O(D)$ readout rounds (right).
(b)~Matching construction. Before readout, the outcomes of a join (dashed) between Bell pairs (arcs) are copied onto the blue sites, $z$ leftward and $x$ rightward, along the near half of each adjacent block and up the arms. Reading the same $\Theta(D^2)$ copies in each of $t\propto D$ rounds (right) gives $\Theta(D^3)$ reports per outcome, $3\times3\times3=27$ as drawn.
(c)~The same number of spare qubits, in a reservoir attached to $A$ (top) or distributed along the route (bottom). The top layout has height-one arms away from the reservoir, so only $O(D)$ sites lie near each distant edge, giving $\ln L=\Theta(D^2)$. Arms of height $H\propto D$ put $\Theta(D^2)$ sites near every interior edge and attain $\ln L=\Theta(D^3)$.}
\label{fig:overview}
\end{figure}

We show that this is impossible when a single route connects the endpoints, so that entangling them requires gates across every edge of the route. On a square lattice with a bounded number of qubits per site, every depth-$D$ protocol starting from a product state leaves endpoints more than $\exp[O(D^3)]$ steps apart exactly separable in every record. This is because only measurements within $D$ steps of an edge, inside its light cone [Fig.~\ref{fig:overview}(a)], can observe what crossed it. An instantaneous broadcast adds no evidence, since it gives every recipient a copy of the same report. At most $O(D^2)$ sites lie that close to an edge, and each supplies at most $O(D)$ reports. Even in the most convincing record, the contribution in which every report near a given interior edge comes from an independent fair coin has relative weight at least $[p/(1-p)]^{O(D^3)}$. With enough disjoint cuts along the route, every record leaves the endpoints separable.

The bound $\exp[O(D^3)]$ is tight. We attain it with a nondestructive measurement of the parity of two remote qubits, a basic primitive of lattice surgery~\cite{Horsman2012,Chamberland2022}. Before readout, we copy each outcome onto about $D\times D$ spare qubits within $D$ steps of its join, along the route and up transverse arms [Fig.~\ref{fig:overview}(b)], and read each copy about $D$ times. Majority votes over these reports, of order $D^3$ per outcome, maintain a fixed accuracy up to distances $\exp[\Theta(D^3)]$, without discarding any run. If the same number of spare qubits is instead concentrated in a reservoir at one end [Fig.~\ref{fig:overview}(c)], most joins lie more than $D$ steps away from it, and the reach drops to $\exp[\Theta(D^2)]$. A similar count shows that, even with many routes, entangling two qubits a distance $L$ apart in constant depth requires at least of order $\ln L$ qubits per unit distance.

Earlier work bounds the depth savings from feedforward in terms of the number of measured regions~\cite{Friedman2023}, and the entanglement sustained by a repeater chain in terms of the number of parallel uses of a noisy channel~\cite{Masajada2026}. Trinh \emph{et al.}\ proved a finite entangling distance for chains whose links all carry the same full-rank two-qubit state, even under adaptive postselection~\cite{Trinh2026}. Here the circuit can prepare pure links and spend part of its depth copying and rereading the outcomes of the joins. How far it can entangle then depends on how many independent reports it can gather near each cut.

\smallskip
\paragraph*{Setting.}
We model the device as an interaction graph whose vertices, called sites, each hold at most a fixed number of qubits. A layer of gates acts on disjoint supports, each within a site or across two neighboring sites, and in each readout round every qubit can be read at most once by nondemolition measurement. All qubits except the two data qubits start in a product state across sites. A route is a path of neighboring sites between the two endpoint sites, and a join is a Bell measurement on the adjacent ends of two Bell pairs. Each binary measurement updates the quantum state according to its true outcome, but the controller receives a report that is flipped with probability $0<p<1/2$, independently across qubits and rounds. The reports of a run form its record. The controller may choose later operations based on earlier reports and may postselect on any set of records. The depth $D$ counts all gate layers and readout rounds up to the final correction. We take classical processing and communication to be instantaneous, so $D$ excludes classical latency. Our constructions assume ideal gates and memories, and the upper bounds on reach also hold with additional local noise on gates and idle qubits.

As a benchmark task we take a nondestructive measurement of the parity $Z_AZ_B$ of two remote data qubits. The measurement must return the parity and preserve every superposition within the reported sector, which reading out the two qubits separately in $Z$ would destroy. Its operation error $\eps_{\rm op}$ is half the diamond-norm distance to the ideal measurement, including the returned bit and allowing inputs entangled with a reference. An accurate implementation must create entanglement. Applied to $\ket{++}$, the ideal measurement followed by a flip of $B$ on odd parity produces $\Phi^+$, so an implementation with error $\eps_{\rm op}$ leaves the corrected pair with average fidelity at least $1-\eps_{\rm op}$ to $\Phi^+$. Since a separable pair has fidelity at most $1/2$, achieving $\eps_{\rm op}<1/2$ requires the endpoints to be entangled in at least one record. As we will see, returning the correct parity is the easier half of the task.

\begin{figure}[t]
\centering
\includegraphics[width=\columnwidth]{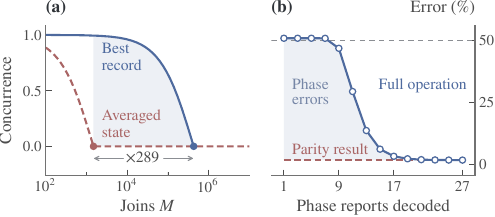}
\caption{\textbf{What averages and parity outcomes miss.}
(a)~Endpoint concurrence~\cite{Wootters1998} for the chain of the introduction ($M$ joins, three reports per outcome, $p=1\%$). The averaged state, decoded by majority vote, becomes separable at 1479 joins, whereas the best record, in which every triple is unanimous, stays entangled up to $427\,597$ joins, 289 times farther (shaded interval). Such records become increasingly rare as the chain grows.
(b)~A fixed circuit prepares the long-range Bell pair before coupling to the data ($L=8399$, $D=20$, $p=10\%$) and acquires 27 reports per outcome at each join, as drawn in Fig.~\ref{fig:overview}(b). Bit corrections use all of them, phase corrections only the number on the horizontal axis. The error of the returned parity (red dashed) stays fixed. The shaded gap to the full operation error (circles) is the probability of returning the correct parity with a phase flip on the data. The gray line marks $\eps_{\rm op}=1/2$.}
\label{fig:finite}
\end{figure}

\smallskip
\paragraph*{Main result.}
Averaging over runs can hide entanglement that individual records retain. With three reports per outcome, the endpoint pair in the most favorable record remains entangled across nearly 300 times as many joins as the averaged pair [Fig.~\ref{fig:finite}(a)]. We therefore define two distances, one for how far some record can remain entangled and one for how far the parity measurement works reliably. At fixed local qubit capacity, both reaches are optimized over adaptive protocols and single-route layouts. These layouts are nearest-neighbor subgraphs of the square lattice with a unique simple path of length $L$ between the endpoint sites. The \emph{entangling reach} $L_{\rm ent}(D)$, which we also call the horizon, is the largest $L$ at which some protocol of depth at most $D$, starting from a product state, leaves the endpoints entangled in at least one record of nonzero probability. The \emph{reliable reach} $L_\eps(D)$ is the largest $L$ at which the parity measurement can be implemented with $\eps_{\rm op}\le\eps$ without discarding any run. Because an error below $1/2$ requires entanglement in some record, $L_\eps\le L_{\rm ent}$ whenever $\eps<1/2$.

The two reaches differ in how the reports are used. With postselection, the controller can wait for the most favorable records, in which all reports of each outcome agree. Each agreeing report then adds $\lambda_p=\ln[(1-p)/p]$ to the log-likelihood ratio in favor of the reported value. When every run counts, majority voting on independent reports has an error exponent given by the Chernoff information $\kappa_p=-\ln[2\sqrt{p(1-p)}]$. The ratio $\lambda_p/\kappa_p$ tends to two as $p\to0$ and diverges as $p\to1/2$, where postselection helps most.

\begin{theorem}[Entangling reach at fixed depth]
\label{thm:time}
Consider single-route layouts with at most a fixed number of qubits per site (at least two), and fix $0<p<1/2$ and $0<\eps<1/2$. At any fixed depth, protocols starting from a product state across sites leave the endpoints exactly separable in every record beyond a finite distance, even after local filtering at the endpoints. For all sufficiently large $D$,
\begin{equation}
 \begin{aligned}
 \ln L_{\rm ent}(D)&=\Theta(\lambda_p D^3),\\
 \ln L_\eps(D)&=\Theta(\kappa_p D^3).
 \end{aligned}
 \label{eq:optimal-reach}
\end{equation}
\end{theorem}

The comparison constants depend only on the local qubit capacity and, for the reliable reach, on $\eps$. The depth at which the asymptotic bounds apply can depend on $p$ and $\eps$. For both reaches the cubic scaling comes from the number of reports available near a cut. Proofs are given in the End Matter and the Supplemental Material (SM)~\cite{SM}.

\smallskip
\paragraph*{Proof sketch.}
The argument must hold record by record, because a rare record carries little weight in the average. Steps (i)--(iii) show that in every record the endpoint state is close to separable, and step (iv) makes this exact. The End Matter follows the same steps.
\begin{enumerate}[label=(\roman*),leftmargin=*,itemsep=1pt,topsep=2pt]
\item \emph{A fair-coin component in every report.} A report that is wrong with probability $p$ can be generated by returning the true outcome with probability $1-2p$ and a fair coin toss otherwise. Whichever bit is reported, the coin accounts for at least a fraction $p/(1-p)=e^{-\lambda_p}$ of its probability. This holds report by report, whatever the controller later does, so the unnormalized endpoint state of every record splits into contributions labeled by which reports were coins.
\item \emph{Deleting the gates across a cut.} Fix the complete reported record and the controller's random seed. Take an edge more than $D$ steps from both endpoints and consider the sites within $D$ steps of it [Fig.~\ref{fig:overview}(a)]. In the contribution where every report from the neighborhood is a coin, the measurements there are effectively unread and act as trace-preserving operations. The gates across the edge can then be deleted without changing the endpoint state or the probability of the record, because their light cone contains neither the endpoints nor any measurement outside the neighborhood~\cite{Friedman2023}. Afterwards the two sides evolve separately from a product state, so this contribution is separable.
\item \emph{Successive comparison over disjoint neighborhoods.} For $L\gg D$, choose $K=\Theta(L/D)$ interior cuts spaced $2D+2$ edges apart, clear of the endpoints and the radius-$2D$ region around $A$. Their neighborhoods are disjoint even where the route folds back, since any shortcut would create a second route. If each neighborhood supplies at most $n$ reports, its all-coin contribution has relative weight at least $e^{-\lambda_p n}$ by step~(i). Conditioned on a record, the all-coin events of different neighborhoods need not be independent, so we apply the comparison to one neighborhood after another. This bounds the remainder, in which no neighborhood is all coins, by a fraction $\exp[-K e^{-\lambda_p n}]$ of the whole.
\item \emph{Exact separability from an anchor.} Closeness to the separable set is not enough, since any admixture of $\Phi^+$ to $\ket{00}$ is entangled. We therefore bound the remainder by an anchor, a product contribution of the same state. Let $n_0$ bound the number of reports within $2D$ steps of endpoint $A$. Replacing all those reports by coins gives a product contribution of relative weight at least $e^{-\lambda_p n_0}$. Its factor at $A$ does not change under further conditioning on coins in distant neighborhoods, whose light cones never meet that of $A$. Step~(iii) bounds the remainder in operator order by a multiple of this anchor that shrinks with $L$. A local rescaling turns the anchor into the identity on its support, and by a separability lemma related to Refs.~\cite{Gurvits2002,Hanson2020} (Lemma~\ref{lem:one-sided-separable}) the identity absorbs any positive perturbation up to half its size. The endpoints are then exactly separable in every record of nonzero probability whenever
\begin{equation}
 K e^{-\lambda_p n}\ge \lambda_p n_0+\ln 3.
 \label{eq:criterion-main}
\end{equation}
\end{enumerate}
On the square lattice both report budgets are $O(D^3)$. At fixed $p$ and large $D$, Eq.~\eqref{eq:criterion-main} thus gives separability once $\ln L$ exceeds a constant times $\lambda_p D^3$. On a chain the budgets are $O(D^2)$, giving the quadratic exponent. For the upper bound on the reliable reach, we instead replace the reports near a single, randomly chosen edge by coins, without telling the controller which edge. The modified protocol leaves the endpoints separable. A second-moment estimate shows that the two protocols become indistinguishable as the route grows at fixed depth and gives the $\kappa_p$ dependence of the reliable reach~\cite{SM}.

\smallskip
\paragraph*{Matching construction.}
To attain the bound, we use the sites within reach of each join to gather independent reports of its two outcome bits. We divide the route into blocks, prepare a Bell pair across each block, and join neighboring pairs. Each join is a Bell-basis rotation whose readout is postponed, so its two outcomes are stored in the computational basis of the two qubits at the join. Before any readout, CNOT gates copy each stored value along half of its block and up the attached arms [Fig.~\ref{fig:overview}(b)]. The copies stay entangled with the endpoints until the first readout fixes their common value. Every further readout then reports that value with a fresh, independent error. Majority votes determine the outcomes, and their parities give the two corrections.

The three stages compete for the same depth. With half-blocks of $\ell$ sites, arms filled to height $h$, and $t$ readout rounds, preparing and copying along the route takes about $2\ell$ layers, copying up the arms $h$ more, and reading $t$ more, while each outcome receives about $(\ell+1)(h+1)t$ reports. This product is largest when the three stages take equal shares of the depth [SM Fig.~\ref{fig:allocation}(b)], so order $D^3$ reports fit within depth $D$. A majority vote over them errs with probability exponentially small in $\kappa_p D^3$, so exponentially many joins can be protected at once and the reliable reach is attained. For the entangling reach, we leave the endpoints unmeasured and keep only runs in which all reports of each outcome agree. Each accepted outcome is then wrong with probability exponentially small in $\lambda_p D^3$.

\smallskip
\paragraph*{Remote parity measurement.}
We now use the protected pair for the full parity measurement, which must also preserve the coherence that parity statistics do not test. A CNOT from each data qubit onto its half of the pair, followed by $Z$ readout of both halves, returns the parity as the XOR of the decoded values, adjusted for the bit correction of the pair. The data states $\ket{00}$ and $\ket{11}$ leave the pair in $\Phi^+$, and both odd-parity states map it to $\Psi^+$, so the readout reveals the parity and nothing else, preserving superpositions within each sector.

A wrong bit correction reverses the returned parity and shows up in parity statistics. A wrong phase correction instead flips the relative sign within a sector and leaves every parity statistic unchanged. If the pair is an equal mixture of $\Phi^+$ and $\Phi^-$, the circuit returns ideal parity statistics on every input, yet on $\ket{++}$ the parity-dependent correction leaves only an incoherent mixture of $\ket{00}$ and $\ket{11}$. The phase correction comes from the Bell measurements at the joins, since repeating the $Z_AZ_B$ measurement cannot distinguish $\Phi^+$ from $\Phi^-$. In this circuit the two residual errors are independent, and on input $\ket{++}$ their four combinations leave the corrected pair in orthogonal Bell states. Its correlations therefore determine the operation error,
\begin{equation}
 1-\eps_{\rm op}
 =\frac{1+\langle Z_A Z_B\rangle_{\rm corr}}{2}\,
  \frac{1+\langle X_A X_B\rangle_{\rm corr}}{2}.
 \label{eq:parity-phase-main}
\end{equation}
The first factor gives the probability of a correct parity bit and the second the probability of no phase flip~\cite{SM}. If the decoder uses every bit report but only one phase report per join, the returned parity is wrong in only $1.7\%$ of runs, while the operation error is $51\%$ [Fig.~\ref{fig:finite}(b)].

\begin{figure}[t]
\centering
\includegraphics[width=\columnwidth]{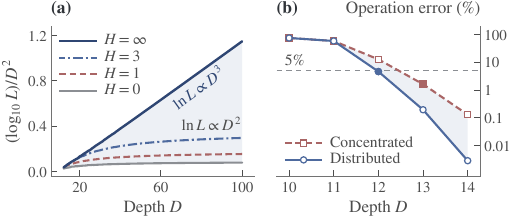}
\caption{\textbf{Reach and accuracy at fixed depth.}
(a)~Reach attained at $p=1\%$ and $\eps=5\%$, plotted as $(\log_{10}L)/D^2$, for arm heights $H$, with the Bell pair prepared before the data arrive. A plateau means that $\ln L$ grows as $D^2$, a linear rise that it grows as $D^3$.
(b)~Minimum operation error at $L=8399$ and $p=10\%$ for the concentrated and distributed layouts of Fig.~\ref{fig:overview}(c), each with $33\,600$ qubits besides the data, optimized over schedules that couple the data during the joins and read all copies after copying. SWAP counts as one gate layer. Filled markers show the first depth below $5\%$ error for each layout.}
\label{fig:reach}
\end{figure}

\smallskip
\paragraph*{Placement of spare qubits.}
For a comb with arms of height $H$, the optimal reliable reach at fixed $p$, $\eps$, and local qubit capacity satisfies
\begin{equation}
 \ln L_\eps(D,H)=\Theta\!\left(D^2\min\{D,H+1\}\right).
 \label{eq:height-reach-main}
\end{equation}
At fixed $H$ the logarithmic reach grows only quadratically, whereas arms of height proportional to $D$ attain cubic growth [Fig.~\ref{fig:reach}(a)]. Extending them farther leaves their tips outside the light cone of the join they serve. Any component attached to a single route meets it at only one site, since a second contact would create a second route. A reservoir at $A$ is therefore out of range of every distant join, no matter how many qubits it holds. At $L=8399$ and $p=10\%$, with the same number of qubits and the same depth, the 12-layer schedules of Fig.~\ref{fig:reach}(b) reach operation errors of $4.65\%$ with distributed arms and $12.74\%$ with a reservoir at $A$. A local circuit without measurements, by comparison, would need at least 4200 layers merely to entangle the endpoints at this distance~\cite{BHV2006,FossFeig2023}.

\smallskip
\paragraph*{Multiple routes.}
When several routes connect the endpoints, some edges can be bypassed. Checks across routes can protect cat-state order~\cite{Zhu2023,Chen2025}, and Bell pairs can be distributed below a noise threshold~\cite{Raussendorf2005,Perseguers2008,Choe2024,Harley2026}. Edges that every route must cross, however, remain as limiting as before. Square-lattice modules joined in series by single links~\cite{Monroe2014,Bravyi2022modular,Main2025}, for example, cannot pass entanglement across more than $\exp[O(D^3)]$ such links, even if each module contains many internal routes~\cite{SM}.

A cost in qubits survives even with multiple routes. Group the sites into levels by their graph distance from one endpoint. One step changes that distance by at most one, so every route to the other endpoint visits every intervening level, and the edges between two consecutive levels form a cut that no route can avoid. A band of $2D+2$ consecutive levels contains such a cut together with its depth-$D$ neighborhood. The roughly $L/2D$ disjoint bands share at most $ND$ reports, where $N$ is the total number of qubits, so a band receives on average at most of order $ND^2/L$ of them. For the endpoints to remain entangled in even one record, the argument above requires this share to grow at least logarithmically with the number of bands. This yields, at any depth~\cite{SM},
\begin{equation}
 ND^2=\Omega_p(L\ln L).
 \label{eq:hardware-main}
\end{equation}
Constant depth therefore requires, on average, at least of order $\ln L$ qubits per unit distance. This parallels the result of Choe and K\"onig that constant-depth Bell-pair generation robust to local stochastic noise on a repeater chain requires $\Omega(\ln L)$ qubits per repeater~\cite{Choe2024}. Narrow strips saturate Eq.~\eqref{eq:hardware-main}. A square-lattice strip of $w\lesssim(\ln L)^{1/3}$ rows contains a comb with arms of height $w-1$, so by Eq.~\eqref{eq:height-reach-main} the parity measurement at fixed accuracy takes depth of order $\sqrt{\ln L/w}$, the least that the bound allows~\cite{SM}. The distributed 12-layer schedule of Fig.~\ref{fig:reach}(b) needs only three rows.

\smallskip
\paragraph*{Discussion and outlook.}
With ideal gates and storage, reliable parity measurement on the best single-route square-lattice layouts needs depth growing only as $(\ln L)^{1/3}$ at fixed readout error and target accuracy. Two powers of $D$ in the logarithmic reach come from the sites available for copies near a join, and the third comes from rereading those copies with independent reporting errors. With readout confined to one final round, spatial copying alone attains the optimal reach $\exp[\Theta(D^2)]$ on a sufficiently tall comb~\cite{SM}.

Even at a readout error of $1\%$, reading each outcome once leaves the endpoints separable after joining just 45 Bell pairs, on the scale of recent superconducting experiments~\cite{Smith2026}. To protect a join, spare qubits must be close enough to receive a copy of its outcome before readout. Benchmarks should also test coherence, since phase flips within a sector leave the returned parity unchanged.

The proof uses independence across reports and a bound on the likelihood ratio of any single reading, and neither can be dropped. Faults that flip reports at neighboring joins in pairs can cancel in the correction parity except at the ends. An ideal Gaussian detector admits sufficiently decisive readings to herald entanglement at any distance, although any fixed finite digitization of its signal restores the horizon~\cite{SM}. Independent phase faults on the prepared Bell links leave the join statistics unchanged while reducing the corrected $XX$ correlation at the endpoints. The converse bounds still hold with additional local gate and storage noise under the same readout model~\cite{SM}.

Multiple routes offer a way around the horizon. Planar strips of polylogarithmic width already support constant-depth Bell-pair generation below a noise threshold~\cite{Harley2026}, but whether the logarithmic width that Eq.~\eqref{eq:hardware-main} requires at constant depth also suffices for reliable long-range operations is open.

\smallskip
\paragraph*{Acknowledgments.}
\begin{acknowledgments}
C.C.\ thanks Jens Eisert for inspiring discussions on quantum measurement and noise.
This work was supported by the Alexander von Humboldt Foundation. The experimental Bell fidelities and readout calibrations are from Figs.~5(a) and~13, respectively, of Ref.~\cite{Smith2026}.
\end{acknowledgments}

\bibliography{refs}

\newpage
\onecolumngrid
\section*{End Matter}
\paragraph*{A criterion for arbitrary interaction graphs.}
We prove the separability criterion behind the four steps of the main text, for unequal report budgets on an arbitrary interaction graph. The subheadings below follow those steps.
Let $d\le D$ count layers that can spread quantum support between sites,
and let $R\le bD$ bound the reports per site, where $b$ is the fixed
local qubit capacity. Initialization is product across sites,
conditional on any initially available classical information.
Let $A$ be one final output qubit and $B$ any collection of output
registers outside the radius-$2d$ region around $A$.
Choose cuts that each separate $A$ from all of $B$. Their radius-$d$
neighborhoods $W_j$ contain every site within graph distance
$d$ of an endpoint of a crossing edge. These neighborhoods must be
mutually disjoint, contain no output in $AB$, and lie outside the
radius-$2d$ region around $A$. Let $n_j$ and $n_0$ be deterministic
report budgets for these neighborhoods and the region around $A$,
respectively, valid for every reported record. Write
$Z_\alpha=\sum_j e^{-\alpha n_j}$ for the sum over interior cuts.
The neighborhood of $A$ is counted separately through $n_0$.

\begin{theorem}[Separability in every record]
\label{thm:range}
Under these conditions, the output on $AB$ is separable in every
record of nonzero probability whenever
\begin{equation}
 Z_{\lambda_p}\ge\lambda_p n_0+\ln3.
 \label{eq:cut-range}
\end{equation}
The complete record may be retained. Record selection and local
filters on $A$ and $B$ preserve the conclusion.
\end{theorem}
Cuts with the smallest report budgets contribute most to $Z_{\lambda_p}$. Each term bounds from below the relative weight of a cut with
uninformative reports. The right side accounts for the product
contribution supplied near $A$, which converts proximity to the
separable set into exact separability [step (iv)]. If all $K$ interior budgets satisfy $n_j\le n$, then
$Z_{\lambda_p}\ge K e^{-\lambda_p n}$, which gives Eq.~\eqref{eq:criterion-main}.

\smallskip
\paragraph*{Step (i): a fair-coin component in every report.}
For a binary on-site instrument $\{\mathcal M_0,\mathcal M_1\}$, whose
sum is trace preserving, the noisy map for report $y$ and its
fair-replacement contribution satisfy
\begin{equation}
 \begin{aligned}
 \mathcal A_y&=(1-p)\mathcal M_y+p\mathcal M_{1-y},\\
 \mathcal U_y&=p(\mathcal M_0+\mathcal M_1),\\
 e^{-\lambda_p}\mathcal A_y&\preceq_{\rm CP}\mathcal U_y
       \preceq_{\rm CP}\mathcal A_y.
 \end{aligned}
 \label{eq:local-order}
\end{equation}
Here $\mathcal X\preceq_{\rm CP}\mathcal Y$ means
$\mathcal Y-\mathcal X$ is completely positive. Subtracting the fair
contribution from $\mathcal A_y$ leaves $(1-2p)\mathcal M_y$.
Subtracting $e^{-\lambda_p}\mathcal A_y$ from $\mathcal U_y$ leaves
$p(1-2p)\mathcal M_{1-y}/(1-p)$.

\smallskip
\paragraph*{Step (ii): deleting the gates across a cut.}
Give each region one independent slot per report allowed by its budget,
indexed by order of use. Assign independent slots to the remaining reports. Each slot replaces the true outcome by a fair bit with
probability $2p$, including unused slots in the auxiliary probability
space. Independence makes this valid under adaptive scheduling.
Fix the reported record $\boldsymbol y$, replacement flags, fair bits, and controller
randomness $z$. Gates, bases, and schedules are fixed.
In a fully replaced cut neighborhood $W_j$,
summing true outcomes leaves trace-preserving maps after scalar report
weights are factored out. All selected,
non-trace-preserving maps lie outside it.

Let $\mathcal K_{\boldsymbol y,z}$ be the resulting map with only $AB$ retained, and
$\widetilde{\mathcal K}_{\boldsymbol y,z}$ the map after deleting every gate across
the associated cut. Backward propagation of an observable $O_{AB}$ gives
\begin{equation}
 \mathcal K_{\boldsymbol y,z}^{\dagger}(O_{AB})
 =\widetilde{\mathcal K}_{\boldsymbol y,z}^{\dagger}(O_{AB}).
 \label{eq:adjoint}
\end{equation}
The adjoints of trace-preserving maps are unital. The supports of $O_{AB}$ and of all
selected, non-trace-preserving maps lie outside $W_j$, at least $d+1$
graph steps from the nearest crossing-edge endpoint. None can reach
a deleted gate in $d$ layers. The equality includes the identity,
so it preserves the probability of the record. After deletion, the initial
state and fixed evolution factor across the cut. The output contribution
is a positive product operator, and averaging over $z$ preserves separability.
Only trace preservation and the supports of the layers enter here, so
gates and idle steps may be replaced by arbitrary local channels.
For transmission, place an untouched reference qubit beside the input.
Separability of the Choi operator of each record then makes the corresponding
receiver map entanglement breaking~\cite{HSR2003}. A related argument for
noisy quantum buses counts cuts where depolarization destroys the
entanglement shared by neighboring stations~\cite{Choe2024}.

\smallskip
\paragraph*{Step (iii): successive comparison over disjoint neighborhoods.}
Let $E_j$ denote full replacement in $W_j$ and $\Gamma$ an event determined
by flags outside $W_j$. For the map $\mathcal N_{\boldsymbol y}$ of a reported record, iteration
of Eq.~\eqref{eq:local-order} gives
\begin{equation}
 \mathcal N_{\boldsymbol y}^{\Gamma\cap E_j}
 \succeq_{\rm CP}e^{-\lambda_p n_j}\mathcal N_{\boldsymbol y}^{\Gamma} .
 \label{eq:window-domination}
\end{equation}
Used slots give the local comparison and unused ones give $2p\ge e^{-\lambda_p}$.
The regional padding makes the exponent $n_j$ valid for any fixed
budget that holds along every record. Applying the comparison
successively to disjoint neighborhoods yields
\begin{equation}
 \begin{gathered}
 \Pi=\prod_j(1-e^{-\lambda_p n_j}),\\
 \mathcal N_{\boldsymbol y}=\mathcal E_{\boldsymbol y}+\mathcal G_{\boldsymbol y},\qquad
 0\preceq_{\rm CP}\mathcal G_{\boldsymbol y}
       \preceq_{\rm CP}\Pi\mathcal N_{\boldsymbol y} ,
 \end{gathered}
 \label{eq:branch-domination}
\end{equation}
where $\mathcal E_{\boldsymbol y}$ is entanglement breaking for transmission.
The corresponding preparation contribution is separable. No
independence of flags conditioned on $\boldsymbol y$ is assumed.

For transmission, retain $\boldsymbol y$ at the receiver and apply a
record-dependent trace-nonincreasing CP filter only there. For
preparation, allow record-dependent local filters on $A$ and $B$.
Summing accepted records preserves both separability and CP order.
Let $\rho_{\rm acc}$ be the normalized accepted state. Its Bell fidelity
$\Fe^{\rm acc}$ is evaluated on the reference and output for transmission,
or on the two endpoints for preparation. Then
\begin{equation}
 \Fe^{\rm acc}\le\frac{1+\Pi}{2},\qquad
 \inf_{\sigma\in{\rm Sep}}\frac12
       \|\rho_{\rm acc}-\sigma\|_1\le\Pi .
 \label{eq:uniform-herald}
\end{equation}
This requires only positive acceptance probability. When it depends
on the input, the normalized accepted operation need not be a channel.

\begin{lemma}[A separable neighborhood]
\label{lem:one-sided-separable}
If $0\le X\le\Id/(\dim A)$, then
$\Id+X$ is separable across $A:B$, independently of the dimension of $B$.
\end{lemma}
\begin{proof}
For normalized Haar measure on pure states of $A$,
\begin{equation}
 \begin{aligned}
 \Id+X&=(\dim A)\int d\psi\,\ket\psi\bra\psi\otimes T_\psi,\\
 T_\psi&=\Id_B-\tr_A X
              +(\dim A+1)\langle\psi|X|\psi\rangle_A\ge0 .
 \end{aligned}
 \label{eq:strict-ball}
\end{equation}
The identity follows from the Haar second moment and positivity
from $\tr_A X\le\Id_B$. The operator-order condition involves only the dimension of $A$.
\end{proof}

\smallskip
\paragraph*{Step (iv): exact separability from an anchor.}
Fix the reported record and the controller's random seed. Write $\omega$
for the unnormalized endpoint state and $\omega_{\Gamma}$ for the contribution
from flag event $\Gamma$. Let $E_0$ denote full replacement in the radius-$2d$
neighborhood of $A$, with report budget $n_0$. Its contribution obeys
\begin{equation}
 e^{-\lambda_p n_0}\omega\le\sigma_0\le\omega,\qquad
 \sigma_0:=\omega_{E_0}=\rho_A\otimes\sigma_B .
 \label{eq:strict-anchor}
\end{equation}
The causal past of $A$ shares no initial site or gate with the pasts
of the distant outputs and outside selected maps. A shared initial site
would require separation at most $2d$. A common gate in interaction layer
$j$ would require separation at most $2(d-j)+1$. The normalized factor
$\rho_A$ stays fixed under conditioning on outside flags because its cone
contains the same nonselective maps fixed by the record.

Let $E=\bigcup_{j\ge1}E_j$. The cut neighborhoods are disjoint from
the region defining $E_0$, so Eq.~\eqref{eq:window-domination} also
applies with $E_0$ fixed. Thus $\omega_{E_0\cap E}\ge(1-\Pi)\sigma_0$.
Both sides have the same $A$ factor, so their difference is a positive
product operator. The contribution from $E\setminus E_0$ is separable
by cut deletion. Grouping these two terms into a separable operator
$\sigma$ gives
\begin{equation}
 \begin{aligned}
 \omega &=(1-\Pi)\sigma_0+\sigma+\omega_{E^c},\\
 0\le\omega_{E^c}&\le\Pi\omega\le\Pi e^{\lambda_p n_0}\sigma_0 .
 \end{aligned}
 \label{eq:strict-choi}
\end{equation}
The cut criterion gives $\Pi\le e^{-Z_{\lambda_p}}\le e^{-\lambda_p n_0}/3$.
If $\sigma_0=0$, then $\omega=0$. Otherwise restrict to its support
and conjugate locally by $\rho_A^{-1/2}\otimes\sigma_B^{-1/2}$, which
sends $\sigma_0$ to the identity. The transformed sum of the first and
last terms is $(1-\Pi)(\Id+X)$ with
$0\le X\le\Pi e^{\lambda_p n_0}\Id/(1-\Pi)\le\Id/2$.
Lemma~\ref{lem:one-sided-separable} applies on this support since its
$A$ dimension is at most two. Undoing the congruence and adding
$\sigma$ proves separability. Averaging controller seeds, selecting
records, and applying endpoint filters preserve it.

\clearpage
\onecolumngrid
\begin{center}
\textbf{\large Supplemental Material for\\
Readout errors limit how far adaptive quantum circuits can entangle}\\[3pt]
\end{center}
\setcounter{equation}{0}
\setcounter{figure}{0}
\setcounter{table}{0}
\setcounter{section}{0}
\renewcommand{\theequation}{S\arabic{equation}}
\renewcommand{\thefigure}{S\arabic{figure}}
\renewcommand{\thetable}{S\arabic{table}}
\renewcommand{\theHtable}{S.\arabic{table}}
\renewcommand{\thesection}{S\arabic{section}}
\renewcommand{\theHequation}{S.\arabic{equation}}
\renewcommand{\theHfigure}{S.\arabic{figure}}
\renewcommand{\theHsection}{S.\arabic{section}}
\setcounter{secnumdepth}{2}

Once the endpoints share a Bell pair, they can teleport an unknown qubit or use the two halves as ancillas for a nondestructive parity measurement (Sec.~\ref{sm:operations}). Conversely, transmission must preserve entanglement with a reference, and parity measurement must pass the corrected Bell test on $\ket{++}$. Section~\ref{sm:geometry} uses these tests to bound the reach. Sections~\ref{sm:time}--\ref{sm:unanimity} give the chain bounds, the transmission circuit, and the fidelity and yield of postselection. The final three sections cover GHZ preparation, general detectors, and complementary-basis tests.

\FloatBarrier
\section{Remote operations and encoded endpoints}
\label{sm:operations}

\subsection{Nondestructive parity measurements}
The ideal remote parity instrument, including its returned classical bit, is
\begin{equation}
 \mathcal P_{ZZ}(\rho)
 =\sum_{s=0}^1\ket{s}\bra{s}_C\otimes P_s\rho P_s,
 \qquad P_s=\frac{\Id+(-1)^s Z_AZ_B}{2}.
 \label{eq:remote-parity-instrument}
\end{equation}
On $\ket{++}$, correcting the returned bit $s$ with $X_B^s$ gives
$\Phi^+$ for either ideal outcome. The Bell test is a two-outcome
measurement, so an implementation with half-diamond error $\eps_{\rm op}$
leaves the corrected pair, averaged over records, with fidelity
$F_{\rm corr}$ with $\Phi^+$ obeying
\begin{equation}
 1-F_{\rm corr}\le\eps_{\rm op}.
 \label{eq:bell-bridge-main}
\end{equation}
A separable pair has fidelity at most $1/2$ with $\Phi^+$. Thus
Eq.~\eqref{eq:bell-bridge-main} and Theorem~\ref{thm:range} bound the
reach of the parity instrument,
including accepted operations whose normalized Bell test exceeds $1/2$.
This conclusion concerns the product-input test even if acceptance
depends on the input.

\begin{figure}[t]
\centering
\includegraphics[width=0.55\textwidth]{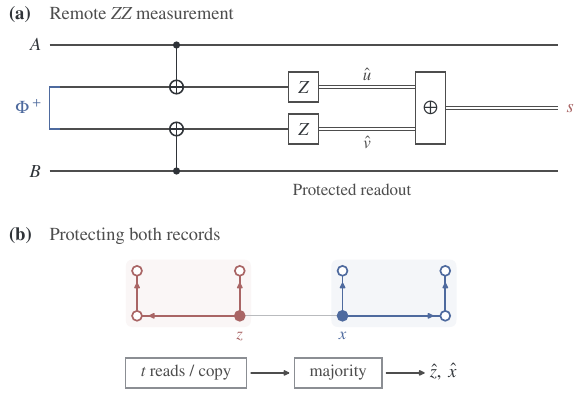}
\caption{\textbf{Protecting the parity result and the data coherence.}
(a) A shared Bell pair and endpoint-local CNOTs implement a nondestructive
$ZZ$ measurement. Protected $Z$ readouts of the resource qubits give decoded values $\hat u,\hat v$ and parity
$s=\hat u\oplus\hat v$. On $|++\rangle$, applying $X_B^s$ should
produce a Bell pair. (b) After the Bell-basis transformation of each join,
the two pointers $z,x$, the qubits that store its outcomes, are copied into disjoint half-blocks and along
their arms. Each value has order $\ell\min\{\ell,H+1\}$ copies,
each read $t$ times, at depth $O(\ell+t)$. Parities of the decoded
values over all joins determine the resource corrections.}
\label{fig:protocol}
\end{figure}

For Fig.~\ref{fig:protocol}(a), prepare a corrected Bell pair on ancillas
$c_A,c_B$, apply $\mathrm{CNOT}_{A\to c_A}$ and $\mathrm{CNOT}_{B\to c_B}$,
then measure the ancillas in $Z$. The data Kraus operator for true
outcomes $u,v$ is $P_{u\oplus v}/\sqrt2$. Returning only $u\oplus v$
implements Eq.~\eqref{eq:remote-parity-instrument}. Related remote gates
also consume shared entanglement~\cite{Eisert2000}.

The resource produced by the protected wire is Bell diagonal. If its
Bell infidelity is $\delta$, its trace distance from $\Phi^+$ is exactly
$\delta$. Contractivity under the fixed instrument processor bounds the
induced half-diamond error by $\delta$, uniformly over data inputs and
spectator references. Read each endpoint pointer, the ancilla holding $u$ or $v$, an odd number $m$ of
times and majority-decode. Each decision has error
$q_m(p)\le e^{-\kappa_p m}$, independently of its true value. A union
bound on the two endpoint errors gives
\begin{equation}
 d_\diamond(\widetilde{\mathcal P}_{ZZ},\mathcal P_{ZZ})
 \le\delta+2q_m(p).
 \label{eq:remote-parity-error}
\end{equation}
Here $d_\diamond$ denotes half the diamond-norm distance between instruments,
including the returned register $C$. Choosing $\delta\le\eps/2$ and
$q_m(p)\le\eps/4$ adds only $O_{p,\eps}(1)$ endpoint layers, using
nondemolition readout or the fresh probes of Sec.~\ref{sm:wire-upper}.

The two resource errors have distinct effects. With resource
$(\Id\otimes X^xZ^z)\ket{\Phi^+}$, the data Kraus operator for true
outcomes $u,v$ is
$K_{uv}^{xz}=(-1)^{zu}Z_A^zP_{u\oplus v\oplus x}/\sqrt2$.
Thus $x$ changes the returned parity and $z$ introduces a phase error
within its sector. An equal mixture of $\Phi^+$ and $\Phi^-$ gives the
ideal parity probabilities for every input, but separable conditional
outputs on $\ket{++}$.
For any Bell-diagonal resource and ideal endpoint reports, the
half-diamond error equals the Bell infidelity. Contractivity gives the
upper bound. On input $\ket{++}$ the correction $X_B^s$ yields four
orthogonal Bell outputs for the four resource errors, attaining the bound. For the wire with
$M$ joins, let $\beta=(1-2q)^M$ denote the parity bias and let each
endpoint decoder err independently with probability $r_{\rm e}$. The same argument
gives the exact full-instrument error
\begin{equation}
 d_\diamond(\widetilde{\mathcal P}_{ZZ},\mathcal P_{ZZ})
 =1-\frac{[1+\beta(1-2r_{\rm e})^2](1+\beta)}4.
 \label{eq:parity-exact-error}
\end{equation}
The output register contains only the decoded parity.

Remote CNOT has the same fixed-accuracy scaling. Its action on
$\ket{+0}$ gives the Bell-test lower bound. For the upper bound, use
$\mathrm{CNOT}_{A\to c_A}$, a $Z$ measurement of $c_A$ and the corresponding
$X_{c_B}$ correction, followed by $\mathrm{CNOT}_{c_B\to B}$, an $X$ measurement
of $c_B$, and the corresponding $Z_A$ correction~\cite{Eisert2000}.
Protect the two reports as above. Preparing the Bell resource before
coupling to unknown data also makes the unanimous-report resource
selection input independent. It supplies accepted parity measurements
and remote CNOTs with the same distance exponents.

These transfers compare tasks at fixed $p$ and fixed accuracy. We use
$\Dopt$ and $\Topt$ for transmission. The uniform detector estimates
below concern this task. For unequal layer times we give a separate law
for parity measurement in Eq.~\eqref{eq:weighted-parity-time}, since the
instrument must acquire its returned bit.

\subsection{Pre-encoded endpoint blocks}
\label{sm:encoded-endpoints}
Allow arbitrary initial encoding within two confined endpoint blocks.
It suffices that the initial state factor across every chosen interior
cut. In particular, the two blocks may contain arbitrary independent
encoded states while the intervening sites start in a product state.
Choose disjoint cut neighborhoods containing no site of either block.
This places every site of a block at least $d+1$ from each crossing-edge endpoint. For a
fully replaced neighborhood, Eq.~\eqref{eq:adjoint} still deletes the
cut, leaving a separable contribution across the endpoint blocks.
Initial entanglement within a block does not affect this factorization.

The comparison in Eq.~\eqref{eq:branch-domination} therefore gives, for
every reported record, $\omega_{\boldsymbol y}=E_{\boldsymbol y}+G_{\boldsymbol y}$ with $E_{\boldsymbol y}$ separable and
$0\le G_{\boldsymbol y}\le\Pi\omega_{\boldsymbol y}$, where
$\Pi=\prod_j(1-e^{-\lambda_p n_j})$.
Any record-dependent joint decoder or trace-nonincreasing local filter
within either endpoint block preserves this comparison and separability.
After selecting and normalizing any positive-probability set of records,
the nonseparable remainder has relative trace at most $\Pi$.
The decoded two-qubit Bell fidelity consequently obeys
\begin{equation}
 F_{\rm log}^{\rm acc}\le
 \frac12\left[1+\prod_j(1-e^{-\lambda_p n_j})\right].
 \label{eq:encoded-endpoints}
\end{equation}

An accurate logical parity measurement on
$\ket{+_{\rm L}}\ket{+_{\rm L}}$, followed by the reported logical
correction, or logical CNOT on $\ket{+_{\rm L}}\ket{0_{\rm L}}$, must
produce logical Bell fidelity at least $1-\eps$.
Since $\Pi\le e^{-Z_{\lambda_p}}$, this requires
\begin{equation}
 \sum_j e^{-\lambda_p n_j}\le -\ln(1-2\eps).
 \label{eq:encoded-report-cost}
\end{equation}
Packing interior cuts with the same chain or comb capacities gives
the corresponding fixed-accuracy lower bounds on depth. The distance
is the gap between the blocks. The bound allows free joint decoding within each block under the
assumed factorization across interior cuts. Encoding and logical
operations may increase the depth for a particular endpoint code.

\paragraph*{Exact separability after logical decoding.}
For a strict statement, assume the initial state has the form
$\rho_{\mathcal A}\otimes\rho_{\mathcal B}\otimes
\bigotimes_{v\notin\mathcal A\cup\mathcal B}\rho_v$,
where $\mathcal A,\mathcal B$ are the full retained endpoint blocks
and $\operatorname{dist}(\mathcal A,\mathcal B)>2d$.
Their internal states are unrestricted. Replace all reports in
$N_{2d}(\mathcal A)$, with budget $n_0$, and choose the separating
cut neighborhoods outside this region and disjoint from both blocks.
At fixed reported record and controller seed, the backward pasts of
$\mathcal A$ and of all distant outputs and outside selected maps share
no gate or initial factor. Neither past reaches the other block's initial state, and the
remaining initial factors are products across sites.
The replaced region gives $\sigma_0=\sigma_{\mathcal A}\otimes\sigma_{\mathcal B}$.
Here $\sigma_{\mathcal A}$ is the normalized output of the nonselective maps in the
block's past for the fixed record and seed. Conditioning on flags
outside the region leaves this factor unchanged. Equations~\eqref{eq:strict-anchor}
and~\eqref{eq:strict-choi} apply to the two blocks.

Apply a record-dependent CP decoder from $\mathcal A$ to one logical
qubit to every term in the decomposition. It preserves CP order,
product structure, and separability, and the decoded anchor factor has
rank at most two. If that factor vanishes, domination makes the whole
decoded state vanish. Otherwise the same support restriction and Haar
identity~\eqref{eq:strict-ball} give exact separability whenever
$\Pi\le e^{-\lambda_p n_0}/3$, hence under Eq.~\eqref{eq:cut-range} in the End Matter.
Any local decoder or filter on $\mathcal B$ preserves the conclusion.
On the encoded product input, the decoded Bell fidelity is therefore
at most $1/2$ in each accepted record.
For a block occupying $u$ consecutive chain sites, the enlarged
anchor has budget $n_0\le R(u+4d)$.

\FloatBarrier
\section{Geometry and ancillary placement}
\label{sm:geometry}
For the main reach law, Secs.~\ref{sec:comb-proof} and~\ref{sec:square-unique-path} bound the number of reports near a cut, and Sec.~\ref{sec:comb-construction} gives the matching circuit. The schedules for Figs.~\ref{fig:reach} and~\ref{fig:finite}(b) are in Sec.~\ref{sec:finite-comb-schedules}. The remaining subsections compare ancillary placements and extend the bounds to strips and trees.

\subsection{Report budgets and refinements of the cut bound}
We use the interaction-depth refinement proved in the End Matter,
with $d\le D$ and at most $R\le bD$ reports per site. The binary
Chernoff information is $\kappa_p=-\ln[2\sqrt{p(1-p)}]$.
For $K$ disjoint cut neighborhoods with budgets $n_j\le n$,
Eq.~\eqref{eq:branch-domination} gives
$\Pi\le\exp(-K e^{-\lambda_p n})$.
By Eq.~\eqref{eq:uniform-herald}, accepted fidelity $1-\eps>1/2$ then requires
$n\lambda_p\ge\ln K-\ln[-\ln(1-2\eps)]$.
The exact product bound also gives $n\lambda_p\ge-\ln(2\eps)>0$.

For $K\ge2$ disjoint neighborhoods with at most $n$ reports each,
fixed-accuracy transmission requires
\begin{equation}
 n\gtrsim_\eps
 \begin{cases}
 \ln K/\kappa_p,&\text{deterministic},\\
 \ln K/\lambda_p,&\text{postselected}.
 \end{cases}
 \label{eq:two-information-costs}
\end{equation}
The constants are independent of $p$. The End Matter gives the
arbitrary-yield bound. We derive the deterministic bound below. The constructions match their
detector dependence on a chain. Majority decoding and unanimous-record selection
have the respective binary discrimination exponents $\kappa_p$ and
$\lambda_p$~\cite{Chernoff1952,DAnjou2021,Regula2024}.
They vanish quadratically and linearly, respectively, as $p\to1/2$.

A protocol may choose the location of a measurement from earlier reports. In adaptive fermionic linear optics, for example, the occupation record determines which mode blocks enter each Bell fusion~\cite{CaoTangEisert2026}.
The proof fixes the cut regions before the run. The budgets in
Theorem~\ref{thm:range} are fixed upper bounds valid for every
measurement record. They may be smaller than $R$ times the number of
sites in a region. Preassign that many independent slots to each
region and consume them in temporal order, breaking ties in a fixed order. Adaptive choices of the measured sites do not
change the reporting law. This justifies the regional padding in the
End Matter without assuming a fixed schedule.

The CP comparison also controls more general events in the auxiliary
replacement flags. Fix a record $\boldsymbol y$ and write its CP contributions
as $\mathcal N_{\boldsymbol y}^z$, with $z_i=1$ denoting replacement at slot $i$.
For every configuration of the other flags, the local inequality gives
$\mathcal N_{\boldsymbol y}^{(1,z_{-i})}\succeq_{\rm CP}
e^{-\lambda_p}[\mathcal N_{\boldsymbol y}^{(0,z_{-i})}+\mathcal N_{\boldsymbol y}^{(1,z_{-i})}]$.
For any event $E$ decreasing in the replacement flags,
\begin{equation}
 \mathcal N_{\boldsymbol y}^E\preceq_{\rm CP}
       \Prb_{\rm iid}(E)\mathcal N_{\boldsymbol y},
 \label{eq:flag-domination}
\end{equation}
where $\Prb_{\rm iid}$ denotes independent replacement with probability
$e^{-\lambda_p}$ at each slot. To prove this without operator-valued conditional probabilities,
apply every Choi-space vector $v$ to define the positive scalar measure
$\mu_v(z)=\langle v|J(\mathcal N_{\boldsymbol y}^z)|v\rangle$.
After normalization, each sequential conditional replacement probability
is at least $e^{-\lambda_p}$. Shared independent uniform random numbers therefore
couple its flags coordinatewise above independent Bernoulli($e^{-\lambda_p}$)
flags. A decreasing event has smaller probability. The resulting
inequality for every $v$ is precisely Eq.~\eqref{eq:flag-domination}.
The zero-mass case is immediate. Unused slots satisfy the same local
comparison since $2p\ge e^{-\lambda_p}$. To obtain a cut bound,
the complementary event must separate the outputs.

\paragraph*{Deterministic comparison.}
Compare the original channel with a mixture in which one uniformly
chosen cut neighborhood supplies only fair reports. We bound their
distance by the second moment of the mixture likelihood ratio, a
method also used in quantum state certification~\cite{ChenLiODonnell2022}.
Pad each of $K$ disjoint neighborhoods to $n$ independent
Bernoulli($p$) error slots. Their product law is $\mu$.
For a fixed error array $e$, the adaptive protocol defines a
trace-preserving channel $\channel_e$, including its complete
reported record. Thus $\channel=\mathbb E_\mu\channel_e$.
Let $\mu_j$ replace the error bits in neighborhood $j$ by independent
fair bits. The used reports there become independent of their true
outcomes, so $\mathcal S_j=\mathbb E_{\mu_j}\channel_e$ is entanglement
breaking. Unused slots integrate out.

The likelihood ratios $\mathcal L_j=d\mu_j/d\mu$ are independent, with
$\mathbb E_\mu\mathcal L_j=1$ and
$\mathbb E_\mu\mathcal L_j^2=[4p(1-p)]^{-n}=e^{2\kappa_p n}$.
For $\mathcal S=K^{-1}\sum_j\mathcal S_j$, the triangle inequality
and $\|\channel_e\|_\diamond=1$ give
\begin{equation}
 \begin{aligned}
 d_\diamond(\channel,\mathcal S)
 &\le\frac12\mathbb E_\mu\left|1-\frac1K\sum_j\mathcal L_j\right|\\
 &\le\frac12\sqrt{\frac{e^{2\kappa_p n}-1}{K}} .
 \end{aligned}
 \label{eq:statistical-eb}
\end{equation}
The comparison is uniform over reference-entangled inputs.
Since $\mathcal S$ is entanglement breaking, fidelity at least
$1-\eps$ requires
\begin{equation}
 2\kappa_p n\ge\ln[1+K(1-2\eps)^2].
 \label{eq:detector-resource}
\end{equation}
This proves the deterministic part of Eq.~\eqref{eq:two-information-costs}.
For unequal budgets the mixing weights can depend on the cut.
The likelihood ratios for separate regions fluctuate independently
with variances $v_j=e^{2\kappa_p n_j}-1$.
For normalized weights $w_j$, Cauchy--Schwarz gives
$(\sum_j w_j^2v_j)(\sum_j v_j^{-1})\ge1$ when all $v_j>0$.
Mixing the entanglement-breaking channels with $w_j\propto 1/v_j$
attains equality and gives
\begin{equation}
 \inf_{\mathcal S\in\mathrm{EB}}
 d_\diamond(\channel,\mathcal S)
 \le\frac{1}{2\sqrt{\displaystyle\sum_j
                      (e^{2\kappa_p n_j}-1)^{-1}}}.
 \label{eq:weighted-statistical-eb}
\end{equation}
If some $n_j=0$, the channel is already entanglement breaking and the
right side is zero. For positive budgets the sum in the denominator is at
least $Z_{2\kappa_p}$, so $1/(2\sqrt{Z_{2\kappa_p}})$ is a simpler upper
bound. Equation~\eqref{eq:weighted-statistical-eb} retains the stronger
finite-size estimate and recovers Eq.~\eqref{eq:statistical-eb} for equal budgets.

Finally, a bounded output region $A$ obeys the same
strict graph criterion with $\ln(\dim A+1)$ in place of $\ln3$.
The anchor is the radius-$2d$ neighborhood of the whole output region.
Lemma~\ref{lem:one-sided-separable} gives separability of
$\Id+X$ for $0\le X\le\Id/(\dim A)$, independently of the receiving
dimension. Thus the End Matter applies with
$\Pi\le e^{-\lambda_p n_0}/(\dim A+1)$. Each cut must separate the entire receiving region.

\subsection{Finite-height combs: exact geometric counts}
\label{sec:comb-proof}

The comb has vertices $(x,y)$ with $0\le x\le L$ and $0\le y\le H$.
Edges join $(x,y)$ to $(x,y+1)$ along each arm and $(x,0)$ to
$(x+1,0)$ along the backbone.  It is a degree-three tree embedded
with nearest-neighbor edges in a square lattice.  The two endpoints
of interest are $(0,0)$ and $(L,0)$.
No horizontal connections between different arms are available.

For a backbone edge the number of sites in its radius-$d$ neighborhood
is at most
\begin{equation}
 \begin{split}
 v_d(H)&=2\sum_{r=0}^{d}\min\{H+1,r+1\}\\
 &=\begin{cases}
 (H+1)(2d-H+2),&H\le d,\\
 (d+1)(d+2),&H\ge d .
 \end{cases}
 \end{split}
 \label{eq:comb-window}
\end{equation}
For the anchor around a backbone site, the corresponding bound is
\begin{equation}
 \begin{split}
 u_d(H)&=\min\{H+1,2d+1\}
             +2\sum_{r=1}^{2d}\min\{H+1,r\}\\
 &=\begin{cases}
 (H+1)(4d-H+1),&H\le2d,\\
 (2d+1)^2,&H\ge2d .
 \end{cases}
 \end{split}
 \label{eq:comb-anchor}
\end{equation}
These are the exact counts for an infinite backbone.  Endpoint
boundaries only reduce them.
Spacing cut edges by $2d+2$ along the backbone gives disjoint windows.
Excluding the anchor and endpoint windows still leaves
$K\ge L/[8(d+1)]$ when $L\ge16(d+1)$.
Consequently, the endpoint output is separable in every reported
record if
\begin{equation}
 L\ge
 \max\!\left\{16(d+1),
 8(d+1)[\lambda_p R u_d(H)+\ln3]\,
                 e^{\lambda_p R v_d(H)}\right\}.
 \label{eq:comb-exact-separability}
\end{equation}

At fixed local reporting capacity, $d\le D$ and $R\le bD$.
The volumes in Eqs.~\eqref{eq:comb-window} and
\eqref{eq:comb-anchor} are of order
$(d+1)\min\{d+1,H+1\}$, uniformly in $H$.
Thus any protocol producing an entangled endpoint pair, or
non-entanglement-breaking transmission, in even one record must satisfy
$\ln L=O_{p,b}(D^2\min\{D,H+1\})$ in the sublinear-depth regime.
If $L<16(d+1)$ the required lower bound is immediate instead.
The resulting bound is
\begin{equation}
 D=\Omega_{p,b}\!\left(
 \max\left\{(\ln L)^{1/3},
       \sqrt{\frac{\ln L}{H+1}}\right\}\right).
 \label{eq:comb-depth-lower}
\end{equation}
It holds at arbitrary positive postselection yield.

\subsection{Optimality among square-lattice layouts with a unique path}
\label{sec:square-unique-path}
\begin{corollary}[Optimal reach with a unique connecting path]
Let $G$ be any nearest-neighbor subgraph of the square lattice, with a
unique simple path of length $L$ between prescribed endpoint sites.
Assume product initialization across sites, fixed local qubit capacity,
and at most $bD$ reports per site in depth $D$. At fixed
$0<p<1/2$, every reported two-qubit endpoint state is separable when
$\ln(L+1)>C_{p,b}(D+1)^3$, for a constant independent of $G$.
The largest entangling reach over this class is therefore
$\exp[\Theta_{p,b}(D^3)]$. Tall combs attain this order, also for
complete parity operations at fixed accuracy.
\end{corollary}

\begin{proof}
Write the unique path as $v_0,\ldots,v_L$. Every path edge is a bridge,
and any connected component off the path meets it at most once.
Otherwise a second simple endpoint path would exist. Graph distances
between path vertices therefore equal their distances along the path,
even when its embedding bends back on itself.

For $D\ge1$, choose edges $(v_k,v_{k+1})$ with
$k=3D+1+j(2D+2)$ and $k\le L-D-2$. Their radius-$D$ neighborhoods
are disjoint, avoid the endpoints, and lie outside the radius-$2D$
neighborhood of $v_0$. When $L\ge16(D+1)$, there are at least
$K\ge L/[8(D+1)]$ such cuts. A radius-$D$ graph ball lies inside a
square-lattice Manhattan ball, which has $1+2D(D+1)$ sites.
Consequently each cut neighborhood has at most $4(D+1)^2$ sites,
and the source neighborhood has at most $8(D+1)^2$ sites.

Put $x=8b\lambda_pD(D+1)^2$. Both $\lambda_p n_j$ and
$\lambda_p n_0$ are bounded by $x$. Theorem~\ref{thm:range} then
implies separability whenever $Ke^{-x}\ge x+\ln3$. In particular,
the explicit sufficient distance
\begin{equation}
 L\ge16(D+1)(x+\ln3)e^x,
 \qquad x=8b\lambda_pD(D+1)^2
 \label{eq:square-unique-path-range}
\end{equation}
has logarithm $O_{p,b}((D+1)^3)$. All bounds hold within every
reported record and survive local endpoint filters. The construction in Sec.~\ref{sec:comb-construction} attains the matching reach.
\end{proof}

For any fixed spatial dimension $\nu\ge1$, the same packing argument
applies to nearest-neighbor subgraphs of $\mathbb Z^\nu$ with a unique
endpoint path. Both a cut neighborhood and the endpoint neighborhood
contain $O_\nu((D+1)^\nu)$ sites, each supplying at most $bD$ reports.
Equation~\eqref{eq:cut-range} thus gives $\ln L=O_{p,b,\nu}(D^{\nu+1})$
for any record with entangled endpoints. To attain this exponent, attach a
transverse $(\nu-1)$-dimensional box of side $\ell$ to each backbone
site, with edges between successive boxes only at their roots. The block
circuit copies each outcome of a join along its own half-block and into
the attached boxes, filling $\Theta_\nu(\ell^\nu)$ sites in $O_\nu(\ell)$
layers. With $\ell$ readout rounds, majority decoding suppresses the
error per value exponentially in $\ell^{\nu+1}$. Taking $\ell$ proportional
to $D$ with a sufficiently small prefactor attains
$\ln L=\Omega_{p,\eps,\nu}(D^{\nu+1})$ for the complete parity
measurement at fixed accuracy. For $\nu=1$ each box is a single site.

\paragraph*{Readout dependence of the two reaches.}
For fixed target error $0<\eps<1/2$, the two reaches in
Theorem~\ref{thm:time} obey
\begin{equation}
 \ln L_\eps(D)=\Theta_{b,\eps}(\kappa_pD^3),\qquad
 \ln L_{\rm ent}(D)=\Theta_b(\lambda_pD^3).
 \label{eq:two-reach-costs}
\end{equation}
The comparison constants are independent of $p$. It suffices that $D$
exceed an absolute constant and $\kappa_pD^3\ge C_\eps\ln(D+1)$ for
sufficiently large $C_\eps$. Every fixed $0<p<1/2$ meets this condition
at large depth.

For the upper bound on the reliable reach, apply the random-cut comparison
of Eq.~\eqref{eq:statistical-eb} to the corrected $\ket{++}$ test.
Each modified circuit gives separable endpoints and hence Bell fidelity
at most $1/2$. The likelihood-ratio estimate for this fixed input gives
$K(1-2\eps)^2\le e^{2\kappa_p n}-1$, with at most
$n=4bD(D+1)^2$ reports per cut neighborhood. The packing bound above
then gives $\ln L\le O_{b,\eps}(\kappa_pD^3+\ln(D+1))$, also when
$L<16(D+1)$. Equation~\eqref{eq:square-unique-path-range} gives the
corresponding upper bound $O_b(\lambda_pD^3+\ln(D+1))$ for endpoint entanglement in any
record. Since $\lambda_p\ge\kappa_p$, the stated condition
absorbs both logarithmic terms.

For achievability choose $\ell=h=t=\lfloor(D-5)/4\rfloor$ in
Eq.~\eqref{eq:comb-fixed-distance-schedule}, with $J$ equal blocks
and $L+1=J(2\ell+2)$. The schedule fits within $D$ layers and gives
$m=\Theta(D^3)$ reports per bit, counting the endpoint bits as well.
Majority decoding has complete parity-measurement error at most
$2Je^{-\kappa_p m}$. Taking $J=\lfloor\eps e^{\kappa_p m}/2\rfloor$
gives the first lower bound. For the second, omit the endpoint CNOTs
and all copying and readout of the resource endpoints, retaining those
two qubits as outputs. Condition on a complete record in which
all reports of each outcome agree. For a uniform Bell bit,
$m$ agreeing reports give posterior error $1/(1+e^{m\lambda_p})\le e^{-m\lambda_p}$.
The corrected pair therefore has infidelity at most $2(J-1)e^{-m\lambda_p}$
on these positive-probability records.
Taking $J=\lfloor e^{\lambda_p m}/8\rfloor$ leaves the pair entangled.
Thus the ratio of the two logarithmic reaches is of order
$\lambda_p/\kappa_p$ in the stated regime. This information ratio tends
to two as $p\to0$ and grows as $4/(1-2p)$ as $p\to1/2$. The geometric
prefactors in Eq.~\eqref{eq:two-reach-costs} are not fixed by these limits.

The hypothesis is a unique simple path. A second, longer route would
already invalidate the bridge argument. Planarity alone also permits
the exponentially growing trees of Sec.~\ref{sec:branching-tree}.

The same argument bounds the number of unavoidable edges even when
the endpoint path is not unique. Let $m$ count the bridges whose
removal separates the two endpoints. Contracting the connected
components left after deleting bridges gives a tree, with these $m$
bridges ordered along its endpoint path. Bridge indices differing by
$2D+2$ have endpoint sets at graph distance at least $2D+1$.
Omitting the first $3D+1$ and last $D+1$ bridges leaves the same
packing bound $K\ge m/[8(D+1)]$ when $m\ge16(D+1)$.
Equation~\eqref{eq:square-unique-path-range} therefore remains a
sufficient separability condition with $m$ in place of $L$.
This covers large two-dimensional modules joined in series by single
links, even when each module contains many internal routes.

\paragraph*{A qubit budget on arbitrary interaction graphs.}
Let $N$ be the total number of physical qubits and $L$ the endpoint graph distance.
At fixed $0<p<1/2$, generating an entangled endpoint pair in even one
reported record requires $ND^2=\Omega_p(L\ln(L+1))$. To see this, group
the sites by their graph distance from the first endpoint. A gate crosses
at most one level. When $L\gg D$, the interior levels contain $2k$
disjoint bands of $2D+2$ consecutive levels each, with $k$ of order
$L/(D+1)$. The middle of each band cuts every route to the receiver.

Assign $D$ potential report slots to each qubit. The aggregate band
budget is at most $ND$. Choose the band of smallest budget among the
first $k$ as the anchor and use the last $k$ as separating bands.
Put $x=\lambda_p ND/k$. The anchor budget $n_0$ obeys
$\lambda_p n_0\le x$. For the outer band budgets $n_j$, Jensen's
inequality gives $\sum_j e^{-\lambda_p n_j}\ge k e^{-x}$.

Here the anchor is an interior cut. Fix the actual record and controller
seed. With all reports in the anchor replaced, deleting its crossing
gates leaves a product operator because the initial state factors
across the cut and the frozen maps act separately on its two sides. Conditioning
further on replacement flags in the outer bands changes only its
receiver-side factor, since all those flags lie beyond the anchor.
Thus the same product component and separable-neighborhood argument as in
Eqs.~\eqref{eq:strict-anchor} and~\eqref{eq:strict-choi} applies.
In particular, $k e^{-x}\ge x+\ln3$ suffices for exact separability.
An entangled record therefore requires $x=\Omega(\ln k)$ for large $k$.
For $1\le D\le\sqrt L$, this gives the claimed hardware bound. For
$D>\sqrt L$ it follows already from $N\ge L+1$ on the occupied graph.
At fixed accuracy, the comb of height of order $w$ uses $O(wL)$ qubits
and attains this lower bound when $1\le w\le C[\ln(L+1)]^{1/3}$ for
fixed $C$, by Eq.~\eqref{eq:comb-depth-optimal}.

\subsection{Matching comb construction}
\label{sec:comb-construction}

First construct an endpoint Bell pair with one physical qubit per
comb vertex.  Choose an integer $\ell$ and put
$J=\lfloor(L+1)/(2\ell+2)\rfloor$.
Partition the backbone into $J$ consecutive, disjoint intervals.
The first $J-1$ intervals have $2\ell+2$ vertices, and the last
contains all remaining vertices, between $2\ell+2$ and $4\ell+3$.
Their outer endpoints are the prescribed sites $(0,0)$ and $(L,0)$.
In the asymptotic regime used below, $J\ge2$.

Prepare a Bell pair between the two endpoints of each interval in
parallel.  A local Bell pair followed by SWAP gates moves one half to
the far endpoint in $O(\ell)$ depth, leaving all interior transport
qubits in $|0\rangle$.
The right endpoint of one interval and the left endpoint of the next
are adjacent.  Apply a Bell-basis transformation to each such pair.
Its two qubits now hold computational-basis values specifying the
two eventual Pauli corrections.

Copy the left value leftward through the $\ell$ nearest interior
vertices of its interval, and the right value rightward through the
$\ell$ nearest interior vertices of the following interval.
The two groups occupy disjoint halves, leaving any surplus vertices
in the final interval unused.
Then copy each occupied backbone value upward along its arm to
height $\min\{\ell,H\}$ [Fig.~\ref{fig:protocol}(b)].
Both stages run in parallel across all joins in $O(\ell)$ depth.
Neither final endpoint is copied or measured.
Each Bell value has
$(\ell+1)(\min\{\ell,H\}+1)$ identical computational-basis copies,
including its original pointer.

Omit one copy if needed to leave an odd number, then read each retained copy
an odd number $t$ of times in the same basis. The number $m$ of reports
per Bell value satisfies $m\ge(\ell+1)(\min\{\ell,H\}+1)t/2$, so majority
decoding has error $q\le e^{-\kappa_p m}$.
The exact parity formula, or a union bound over the $2(J-1)$ values,
gives
\begin{equation}
 \begin{split}
 1-F_{\rm pair}&\le2(J-1)e^{-\kappa_p m},\\
 D&=O(\ell+t).
 \end{split}
 \label{eq:comb-explicit-upper}
\end{equation}
Same-basis nondemolition readout gives fresh, independently corrupted
reports of the same stored value.

Take $\ell$ and $t$ of the same order, proportional to
\begin{equation}
 \max\!\left\{
 \left[\frac{\ln(4L/\eps)}{\kappa_p}\right]^{1/3},
 \sqrt{\frac{\ln(4L/\eps)}
              {\kappa_p(H+1)}},\,1\right\}.
 \end{equation}
Then $\kappa_p m\ge\ln(4L/\eps)$ after increasing the fixed
proportionality constant, and the Bell-pair error is at most $\eps/2$.
Integer rounding and the odd-subset convention change only constants.
For fixed $p,\eps$ and sufficiently large $L$, this choice has
$\ell=o(L)$, as required by the interval partition.

For transmission, add an input qubit at the source site and perform a
local Bell measurement onto the prepared pair. A fixed number of rereads
limits the error of its two injection bits to $\eps/2$. This adds
$O_{p,\eps}(1)$ layers and gives Bell fidelity at least $1-\eps$.

For the parity operation, add a data qubit at each endpoint site, which
then holds two qubits, and apply the protected local conversion in Sec.~\ref{sm:operations}.
It has $O_{p,\eps}(1)$ overhead and half-diamond error at most $\eps$.
Combining this construction with
Eq.~\eqref{eq:comb-depth-lower} proves, uniformly over integer $H\ge0$
at fixed $p,\eps$,
\begin{equation}
 D^\star_\eps(L,H)=\Theta_{p,\eps,b}\!\left(
 \max\left\{(\ln L)^{1/3},
       \sqrt{\frac{\ln L}{H+1}}\right\}\right).
 \label{eq:comb-depth-optimal}
\end{equation}
Inverting this bound gives the finite-height reach law stated in the
main text,
\begin{equation}
 \ln L_\eps(D,H)
 =\Theta_{p,\eps,b}\!\left(D^2\min\{D,H+1\}\right).
 \label{eq:comb-reach}
\end{equation}
Leaving the endpoints entangled in even one record
requires this order of depth. The construction uses
$O[L\min\{H+1,D^\star_\eps\}]$ qubits, with $H=0$ giving the chain
and $H$ of order $(\ln L)^{1/3}$ already attaining the long-arm optimum.

\paragraph*{One terminal readout round.}
With a single terminal round, each cut neighborhood has report budget
$O(bd\min\{d,H+1\})$. The same strict bound gives
$d\min\{d,H+1\}=\Omega_{p,b}(\ln L)$ in the shortcut regime.
Taking $t=1$ in the construction attains the resulting depth
$\Theta_{p,\eps,b}(\max\{\sqrt{\ln L},\ln L/(H+1)\})$.
Defer the resource-frame correction and all intermediate pointer
readouts, apply the endpoint CNOTs, and copy the endpoint pointers into
a constant number of unused sites in the first and last blocks.
For true resource frame $(x,z)$ and endpoint results $u,v$, the data
operator is proportional to $Z_A^zP_{u\oplus v\oplus x}$.
After measuring all pointers together, return
$\hat u\oplus\hat v\oplus\hat x$ and apply $Z_A^{\hat z}$.
The hats denote the corresponding decoded values.
This keeps a single terminal readout round, with constant overhead
at fixed $p,\eps$. In particular, sufficiently tall combs give
$\Theta(\sqrt{\ln L})$, whereas the terminal-round chain needs
$\Theta(\ln L)$.

\subsection{Finite-size schedules for the plotted circuits}
\label{sec:finite-comb-schedules}
\paragraph*{Resource prepared before the data arrive.}
In the comb schedules below, $J$ counts blocks and $M=J-1$ counts joins. For the reach calculation use uniform blocks, each with
$2\ell+2$ vertices, so their endpoint distance is $L=(2\ell+2)J-1$.
Choose integers $\ell\ge1$ and $0\le h\le H$. Each Bell value has
$C=(\ell+1)(h+1)$ computational-basis copies, including its original
pointer, and every copy is read $t\ge1$ times. Disjoint copying paths
allow $h>\ell$, with all $h$ layers charged to the schedule. Acquire
and count all $Ct$ reports. For majority decoding keep the largest odd
number $m\le Ct$, omitting one predetermined report when $Ct$ is even. Each endpoint pointer is read an odd
number $m_{\rm e}\ge1$ of times.

Prepare each link on the two central vertices of its block, then move
its halves outward in parallel. This takes one Hadamard layer, one
CNOT layer, and $\ell$ disjoint SWAP layers, leaving all interior
vertices in $|0\rangle$. Table~\ref{tab:comb-layers} gives the remaining
layers for both schedules. Preparing and correcting the resource first gives
\begin{equation}
 D_{\rm sched}=2\ell+h+t+m_{\rm e}+6.
 \label{eq:comb-layer-schedule}
\end{equation}
\begin{table}[tb]
\caption{Parallel layers for a remote parity measurement on equal blocks.
In the staged schedule the endpoint CNOTs run alongside the CNOTs
of the joins. All pointers are read together after copying. At prescribed distance, unequal
block lengths can add SWAP layers as in Eq.~\eqref{eq:comb-fixed-distance-schedule}.}
\label{tab:comb-layers}
\centering
\begin{tabular}{@{}lcc@{}}
\toprule
Stage & Resource first & Staged readout\\
\midrule
Prepare and extend Bell links & $2+\ell$ & $2+\ell$\\
Bell-basis transformations & $2$ & $2$\\
Copy pointers along the backbone and arms & $\ell+h$ & $\ell+h$\\
Read copied pointers & $t$ & $t$\\
Finish at the endpoints & $1+1+m_{\rm e}$ & $1$\\
\bottomrule
\end{tabular}
\end{table}
Each SWAP costs one gate layer. The first schedule corrects the Bell
resource before the endpoint CNOTs and readouts. The staged schedule
leaves only the phase correction after the last measurement.

Let $q=q_m(p)$ be the exact majority error for the $m$ retained
reports, and $r_{\rm e}=q_{m_{\rm e}}(p)$ the endpoint error. Write
$\zeta=(1-2r_{\rm e})^2$ and $\beta=(1-2q)^{J-1}$, since the resource
has $J-1$ joins. Equation~\eqref{eq:parity-exact-error}
gives the full-instrument error $1-(1+\beta\zeta)(1+\beta)/4$.
For $0<\eps<1/2$, provided the endpoint error floor
$(1-\zeta)/2$ is below $\eps$, its largest allowed number of blocks is
\begin{equation}
 \begin{aligned}
 \beta_\ast&=\frac{2(3-4\eps)}{
 \sqrt{(1+\zeta)^2+4\zeta(3-4\eps)}+1+\zeta},\\
 J_{\max}&=1+\left\lfloor
 \frac{\ln\beta_\ast}{\ln(1-2q)}\right\rfloor.
 \end{aligned}
 \label{eq:comb-finite-reach}
\end{equation}
For Fig.~\ref{fig:reach}(a), fix $p=0.01$, $\eps=0.05$, and maximize
$(2\ell+2)J_{\max}-1$ over feasible $\ell,h,t,m_{\rm e}$ with
$D_{\rm sched}\le D$ and $J_{\max}\ge2$. At fixed $\ell,h,m_{\rm e}$
all remaining layers can be assigned to rereads. For large distances we sum the binomial tail by a positive-term recurrence and evaluate
$\ln(1-2q)$ directly from $q$ to avoid rounding $1-2q$ to one.
The $H=0,1$ schedules may both be run on the concentrated layout.

\begin{figure}[t]
\centering
\includegraphics[width=5.2in]{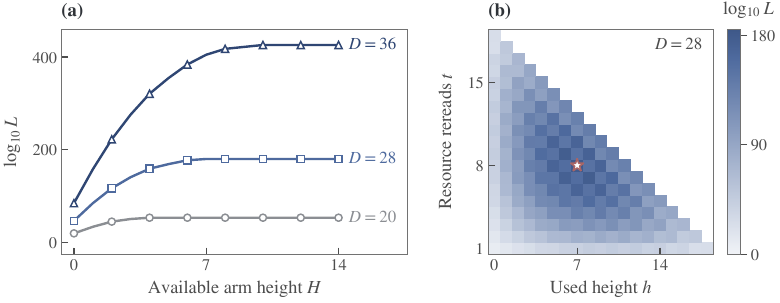}
\caption{\textbf{Arm height and readout rounds at fixed depth.}
(a) Achievable reach versus available arm height $H$ at three fixed
budgets. Taller unused arms do not improve the construction once the
best allocation saturates. (b) At $D=28$, each point fixes the used arm height $h$ and the
number $t$ of readout rounds per resource copy, while optimizing
block length and endpoint readouts. The star marks a maximum, and blank cells mark infeasible allocations. Both
panels assume ideal gates and storage, $p=1\%$, and operation error at most $5\%$.
Every quantum gate and readout layer is included.}
\label{fig:allocation}
\end{figure}

Figure~\ref{fig:allocation}(a) varies $H$ from zero to 14 at
$D=20,28,36$. Panel (b) fixes $D=28$ and optimizes $\ell,m_{\rm e}$
at each feasible $(h,t)$. With fixed endpoint cost and continuous
parameters the report count is maximized at $2(\ell+1)=h+1=t$.
Plotted values use the exact integer schedules.

\paragraph*{Staged readout in Fig.~\ref{fig:reach}(b).}
Prepare the local Bell links as above. At each end, apply a CNOT
from the data qubit to the outer endpoint of its link, in the same layer
as the CNOTs of the joins. All these pairs
are disjoint. The following Hadamards act only at the joins.
The first and last blocks each have a free half-block for copying the
endpoint pointer. The pointers of the endpoints and of the joins can therefore
be copied in the same $\ell+h$ layers and measured in the same $t$
rounds. Each has $(\ell+1)(h+1)$ copies, including the original
pointer. Let $m$ be the largest odd integer at most $(\ell+1)(h+1)t$.
If one report must be omitted, its position is chosen before the run.

Condition on the true outcomes of the joins and denote their correction
parities by $x,z$. The projections at the joins commute with the endpoint
CNOTs, so for true endpoint bits $u,v$ the data operator is proportional
to $Z_A^z P_{u\oplus v\oplus x}$, as in Sec.~\ref{sm:operations}.
Return $\hat u\oplus\hat v\oplus\hat x$ and apply $Z_A^{\hat z}$.
The final phase correction costs one local gate layer. All record-dependent
corrections are thus deferred to a single final feedforward stage, whose
latency on hardware can grow with distance. The depth counts only quantum layers.
The operator
identity above holds for every input, including data entangled with a reference.

For a block size $B=2\ell+2$, let $J=\lfloor(L+1)/B\rfloor$ be the number
of blocks and distribute the $r=L+1-BJ$ excess sites as evenly as
possible among them. Each half-block still fits, and moving the central
Bell pairs to their ends costs at most $\lceil r/(2J)\rceil$ additional
SWAP layers. Counting the Hadamard and CNOT that prepare each local link,
the two layers of the joins, and the final Pauli correction gives
\begin{equation}
 D_{\rm sched}=2\ell+h+t+5+\left\lceil\frac{r}{2J}\right\rceil .
 \label{eq:comb-fixed-distance-schedule}
\end{equation}
Each of the $J-1$ joins contributes one decoded bit to each
correction. The two endpoint decisions also enter the returned parity.
Their majority errors are independent, even when the true endpoint
values are correlated with the data, because the probability of a
majority error is the same for either true bit. With single-bit bias $v=1-2q_m(p)$,
Eq.~\eqref{eq:parity-exact-error} gives
\begin{equation}
 \eps_{\rm op}=1-\frac{(1+v^{J-1})(1+v^{J+1})}{4}.
\end{equation}
At $L=8399$ and $p=0.10$, we minimize this error for each layer budget
$D$ over integer schedules with $D_{\rm sched}\le D$,
$\ell,t\ge1$, $0\le h\le H$, and $J\ge2$.

For the four schedules below, $J=2100$, $r=0$, and $\ell=1$. The
layouts have $33\,600$ resource sites and two endpoint data qubits each.
$H$ is the available arm height along the backbone, $h$ the height
used, and $Ct$ the number of acquired reports per pointer. All acquired
reports count toward the depth, including the one omitted when $Ct$ is even.
\begin{center}
\begin{tabular}{@{}lrrrrrrr@{}}
\toprule
Layout & $D$ & $H$ & $h$ & $t$ & $Ct$ & $m$ & $\eps_{\rm op}$ (\%)\\
\midrule
Distributed & 12 & 3 & 2 & 3 & 18 & 17 & 4.6458\\
Concentrated & 12 & 1 & 1 & 4 & 16 & 15 & 12.7373\\
Distributed & 13 & 3 & 2 & 4 & 24 & 23 & 0.1961\\
Concentrated & 13 & 1 & 1 & 5 & 20 & 19 & 1.6303\\
\bottomrule
\end{tabular}
\end{center}
The distributed 12-layer schedule fits on a $3\times8400$ strip with $25\,202$ qubits including the two data qubits, since it leaves the highest site of each arm idle. The concentrated schedules leave the source reservoir idle. At $p=1\%$, the same distributed 12-layer circuit ($m=17$, $J=2100$) gives operation error $9.50\times10^{-11}$. Reoptimizing within the staged family at this lower noise rate, both layouts first meet the $5\%$ target in ten layers, with $\ell=h=1$, $t=2$ and error $0.143\%$. At $L=8399$,
disjoint backward cones exclude corrected Bell fidelity above $1/2$
for any measurement-free depth below 4200, even after endpoint-local
detection and classical processing.

One can isolate the effect of imperfect links in the same finite schedule.
Suppose each of the $J$ Bell links independently suffers a $Z$ fault on
one half after preparation, with probability $g$. All other quantum
operations are ideal and the binary reporting errors are as above. The fault changes the phase frame but leaves the Bell-outcome
probabilities unchanged. Its contribution to the final phase correction is
the parity of the $J$ faults, so the corrected $XX$ correlation acquires
the factor $(1-2g)^J$. In the staged schedule the operation error is
then $1-(1+v^{J+1})[1+v^{J-1}(1-2g)^J]/4$, with $v=1-2q_m(p)$ as above.
For the distributed 12-layer row, a phase-fault probability of $10^{-5}$ per prepared
link raises the error from $4.65\%$ to $6.56\%$.

\paragraph*{Interleaved copying and readout.}
Suppose a layer can perform gates and nondemolition readouts on disjoint
qubits in the same time. Keep the layer budget of
Eq.~\eqref{eq:comb-fixed-distance-schedule}. As the copies spread, read
each occupied pointer whenever it is idle. During the $\ell$ horizontal
copying layers there are successively $0,\ldots,\ell-1$ such pointers
per value. During vertical copying there are $(\ell+1)j$ in layer $j+1$,
for $0\le j<h$. Copying the bit and endpoint pointers starts one layer
earlier, during the Hadamards of the joins. Their full groups of copies
therefore become available one layer earlier, adding a complete readout
round. A second complete round fits in the final phase-correction layer.
That correction uses the phase reports from completed layers, while the
concurrent bit reports enter only the returned classical parity. With
$C=(\ell+1)(h+1)$ and $t\ge1$ full readout rounds for the phase pointers,
the report counts per value are
\begin{equation}
 n_z=Ct+\frac{\ell(\ell-1)}{2}+(\ell+1)\frac{h(h-1)}{2},
 \qquad n_x=n_z+2C .
\end{equation}
The early $Z$ readouts act only on established copies. Their projectors
commute with subsequent copying, in which these qubits serve only as
controls. Since repeated $Z$ projections on the same value agree, the
data operator for true outcomes remains proportional to
$Z_A^z P_{u\oplus v\oplus x}$, also on reference-entangled inputs.
Round $n_x,n_z$ down to odd counts $m_x,m_z$, discarding any surplus
report at a predetermined position. With $v_x=1-2q_{m_x}(p)$ and
$v_z=1-2q_{m_z}(p)$, the operation error is $1-(1+v_x^{J+1})(1+v_z^{J-1})/4$.

At the same $L=8399$ and $p=0.10$, the following choices illustrate the
gain within the interleaved family. The three rows with $\ell=1$ use $J=2100$ blocks, and the row with $\ell=2$ uses $J=1400$. Each concentrated schedule again
leaves the source reservoir idle.
\begin{center}
\begin{tabular}{@{}lrrrrrrr@{}}
\toprule
Layout & $D$ & $\ell$ & $h$ & $t$ & $m_x$ & $m_z$ & $\eps_{\rm op}$ (\%)\\
\midrule
Distributed & 12 & 1 & 3 & 2 & 37 & 21 & 0.283\\
Concentrated & 12 & 1 & 1 & 4 & 23 & 15 & 6.67\\
Distributed & 13 & 1 & 3 & 3 & 45 & 29 & 0.00412\\
Concentrated & 13 & 2 & 1 & 3 & 31 & 19 & 0.548\\
\bottomrule
\end{tabular}
\end{center}
For the first row, the same prepared-link faults multiply $v_z^{J-1}$
by $(1-2g)^J$, raising the error from $0.283\%$ to $2.33\%$ at
$g=10^{-5}$. For equal blocks and unrestricted arm height, the leading
term of $n_z$ is $\ell h(D-2\ell-h/2)$. Within this schedule family the maximum is
$D^3/27+O(D^2)$, attained to leading order at $\ell=D/6$, $h=2D/3$
up to bounded offsets that leave a full readout round. Waiting until all
copies are ready instead gives a leading count $\ell ht$, with maximum
$D^3/54+O(D^2)$ under $2\ell+h+t=D-O(1)$.

\paragraph*{Using the phase-correction records in Fig.~\ref{fig:finite}(b).}
This comparison uses the resource-first schedule of
Eq.~\eqref{eq:comb-layer-schedule} and holds the quantum circuit and
every measurement fixed. Choose $\ell=h=2$, $t=3$, and $m_{\rm e}=5$, giving $D=20$,
$J=1400$ blocks at $L=8399$, and 27 acquired reports per Bell value.
At $p=0.10$, use all 27 reports for every bit correction, but only a
predetermined odd number $m_z\in\{1,3,\ldots,27\}$ for each phase
correction. Discard the remaining reports without conditioning on
their values. The two correction-error parities remain independent.
Set $\beta_x=(1-2q_{27})^{1399}$,
$\beta_z=(1-2q_{m_z})^{1399}$, and
$\zeta=(1-2q_5)^2$. The parity-result error $e_s$ and complete
instrument error are exactly
\begin{equation}
 e_s=\frac{1-\beta_x\zeta}{2},\qquad
 \eps_{\rm op}=1-\frac{(1+\beta_x\zeta)(1+\beta_z)}4.
 \label{eq:partial-record-error}
\end{equation}
The parity-result error is input-independent. The corrected Bell input
attains the operation error by Eq.~\eqref{eq:parity-exact-error}. On this
fixed circuit $e_s\simeq1.705\%$, while $\eps_{\rm op}$ falls from
$50.85\%$ at $m_z=1$ to $1.713\%$ at $m_z=27$.

\subsection{The same qubit budget with different placements}
\label{sec:equal-budget}
For integer $H\ge1$, compare the uniform height-$H$ comb with a comb
whose arms all have height one except at the source. Extend the source
arm to height $1+(L+1)(H-1)$. Both graphs have exactly
$(L+1)(H+1)$ qubits, maximum degree three, backbone distance $L$,
and a unique path between the endpoints. A common source input qubit
may be added for transmission, or one common data qubit at each endpoint
for the parity operation.

For the concentrated layout and $d\ge1$, the source anchor contains
at most $6d+1$ sites. The disjoint cut neighborhoods chosen beyond it
contain at most $4d+2$ sites each, independently of the long arm's
height. The packing still gives $K\ge L/[8(d+1)]$.
Theorem~\ref{thm:range}, with $R\le bD$ and $d\le D$, therefore excludes
all endpoint entanglement unless $\ln L=O_{p,b}(D^2)$.
Ignoring the extra source qubits gives the matching deterministic
$\Theta(\sqrt{\ln L})$ construction. With uniform arms of height $H=\Theta((\ln L)^{1/3})$,
Eq.~\eqref{eq:comb-depth-optimal} gives the cubic exponent at the same
qubit count.

Any ancillary graph of fixed maximum degree $\Delta$ attached only at
the source leaves the distant cut neighborhoods unchanged.
The source anchor has at most $\exp[O_\Delta(d)]$ sites. Its report budget
enters the right side of Eq.~\eqref{eq:cut-range}, whereas each distant
window contributes $\exp[-O_p(Rd)]$ to the left. Taking logarithms
again gives $\ln L=O_{p,b,\Delta}(D^2)$, independently of the total reservoir
size.

\subsection{Narrow rectangular strips}
\label{sec:narrow-strips}
Consider the full rectangular nearest-neighbor graph with sites
$(x,y)$, $0\le x\le L$ and $0\le y<w$, at fixed local reporting
capacity $b$. Place the two data qubits at $(0,0)$ and $(L,0)$,
with product ancillas. For fixed $p,\eps$ and $C_0>0$, the optimal
depth of the complete parity instrument satisfies
\begin{equation}
 D^{\star}_{\eps,\mathrm{strip}}(L,w)
 =\Theta_{p,\eps,b,C_0}\!\left(\sqrt{\frac{\ln(L+1)}{w}}\right),
 \qquad 1\le w\le C_0[\ln(L+1)]^{1/3}.
 \label{eq:narrow-strip-time}
\end{equation}
The lower bound already holds for creating an entangled endpoint output
from product initialization in any positive-probability record.
Bell-pair distribution and transmission
have the same optimal order.

There are at most $bw(L+1)$ qubits, so the hardware bound in
Sec.~\ref{sec:square-unique-path} gives the stated lower bound. Alternatively, vertical cuts have radius-$d$
neighborhoods of at most $(2d+2)w$ sites, and the endpoint anchor
contains at most $(2d+1)w$. For the upper bound, retain all vertical edges and only the horizontal
edges along the bottom row. The resulting subgraph is a comb of height $H=w-1$, for
which Eq.~\eqref{eq:comb-depth-optimal} gives
$O_{p,\eps,b}(\max\{(\ln L)^{1/3},\sqrt{\ln L/w}\})$.
The width assumption makes the second term dominant up to a
$C_0$-dependent constant.

\subsection{Identical ancillary networks along a path}
\label{sec:reservoir-profile}

At every vertex of a backbone of length $L$, attach an identical rooted
ancillary graph, identifying only its root with that vertex. Different
attachments have no edges between them. Assume the entire graph has
maximum degree at most a fixed $\Delta$, with bounded local dimension
and reporting capacity as in the main text. Let $N(k)$ count sites in
one attachment within distance $k$ of its root, including the root.
The ancillary graph may depend on $L$. Define
\begin{equation}
 x_L=\min\{k\in\mathbb N:k\ge1,\ k^2N(k)\ge\ln(L+1)\}.
 \label{eq:reservoir-scale}
\end{equation}
At fixed $0<p<1/2$ and $0<\eps<1/2$, the optimal depth of the
complete remote parity instrument is
\begin{equation}
 D^\star_\eps(L)=\Theta_{p,\eps,b,\Delta}(x_L).
 \label{eq:reservoir-time}
\end{equation}
The lower bound already excludes arbitrarily weak endpoint entanglement
in any positive-probability record, starting from a state product across
sites. Constants are uniform over the ancillary graphs with this degree
bound, including graphs whose size and shape vary with $L$.

For a backbone cut, its radius-$d$ neighborhood contains at most
$v_d=2\sum_{r=0}^dN(r)$ sites. The radius-$2d$ region at the endpoint
contains at most $u_d=\sum_{r=0}^{2d}N(r)$ sites.
Cuts spaced by $2d+2$, excluding this region and the opposite endpoint,
supply $K\ge L/[8(d+1)]$ disjoint neighborhoods for
$L\ge16(d+1)$. Theorem~\ref{thm:range} gives separability whenever
\begin{equation}
 L\ge\max\left\{16(d+1),\,
 8(d+1)(\lambda_p Ru_d+\ln3)e^{\lambda_p Rv_d}\right\}.
 \label{eq:reservoir-strict}
\end{equation}
Bounded degree implies $\ln(u_D+1)=O_\Delta(D+1)$.
Using $d\le D$, $R\le bD$, and $v_D\le2(D+1)N(D)$, any entangled
record must therefore satisfy
$\ln(L+1)\le C_{p,b,\Delta}D^2N(D)$.
This also holds for $L<16(D+1)$ after increasing the constant.
For $f(k)=k^2N(k)$, monotonicity gives
$f(k)\ge(k/D)^2f(D)$ when $k\ge D$.
Hence $x_L\le\lceil\sqrt{C_{p,b,\Delta}}D\rceil$, proving the
lower bound without an assumption of regular growth.

For achievability use the backbone blocks of
Sec.~\ref{sec:comb-construction} with $\ell=x_L$.
Copy each outcome of a join into its assigned $\ell+1$ backbone vertices
and then into the radius-$\ell$ ball of each attached graph.
The groups are disjoint. A breadth-first spanning tree copies the
value through each ball in at most $\Delta\ell$ layers, scheduling
the outgoing edges of each parent separately.
Each outcome thus has $(\ell+1)N(\ell)$ identical pointers.
Keep an odd subset containing at least half of them and read each
$t$ times, with $t$ the least odd integer at least $C_{p,\eps}\ell$. Majority decoding bounds the total
resource error by
$2L\exp[-\kappa_p C_{p,\eps}\ell^2N(\ell)/2]\le\eps/2$
for a sufficiently large constant $C_{p,\eps}$.
The protected endpoint conversion in Sec.~\ref{sm:operations}
adds $O_{p,\eps}(1)$ layers and error at most $\eps/2$.
If fewer than two blocks fit, $L=O(\ell)$ and coherent Bell distribution
with the same endpoint conversion suffices.
The total depth is $O_{p,\eps,\Delta}(x_L)$.

For polynomial growth, Eq.~\eqref{eq:reservoir-time} gives the chain and comb reach laws
in the main text. For exponential growth, constant rescalings of depth
change $N(D)$ exponentially, so one must retain them when discussing
reach. In particular, $N(D)=\Theta(2^D)$ gives $v_D=\Theta(2^D)$,
and Eq.~\eqref{eq:reservoir-strict} already bounds
$\ln L$ by $O_{p,b,\Delta}(D2^D)$. The general depth law does not imply
$\ln L=\Theta(D^2N(D))$ at a prescribed depth.

\subsection{Branching trees}
\label{sec:branching-tree}

Let $\Delta\ge3$ bound the degree of a tree and let the two outputs
have distance $L$. Cutting the unique path gives the packing estimate
above, with cut and anchor neighborhoods of sizes
\begin{equation}
 |W_j|\le2\sum_{r=0}^d(\Delta-1)^r,\qquad
 |C|\le1+\Delta\sum_{r=0}^{2d-1}(\Delta-1)^r .
\end{equation}
With $d\le D$, $R\le bD$,
Eq.~\eqref{eq:cut-range} makes every reported endpoint
state separable beyond a distance
$\exp[\exp(O_{p,b,\Delta}(D))]$.
Hence every such tree requires
$D=\Omega_{p,b,\Delta}(\ln\ln L)$ for an entangled endpoint pair
or non-entanglement-breaking transmission in even one record, at any yield.

This order is attainable on degree-three trees.
Prepare Bell pairs on disjoint backbone edges $(v_{2j},v_{2j+1})$
and apply Bell-basis transformations between successive pairs.
Join each measured backbone vertex by a single edge to the root of a
separate full binary tree of $N$ ancillas. Both the backbone vertex and
the new root have degree at most three. Copy the pointer value across
this edge and down the tree in $O(\ln N)$ layers, measure the ancillas
once, and majority decode.
With $N=O_{p,\eps}(\ln L)$, the total Bell-pair error is at most
$\eps$, while $D=O_{p,\eps}(\ln\ln L)$ and the number of qubits is
$O_{p,\eps}(L\ln L)$.
Adding local input injection with protected readout takes constant
additional depth, so transmission also attains the optimal order $\ln\ln L$.

\FloatBarrier
\section{The chain: finite-size bounds and optimal depth}
\label{sm:time}

For a chain, let
\begin{equation}
 s=2d+2,\qquad K=\left\lfloor\frac{L-1}{s}\right\rfloor,
 \qquad \xi=s e^{sR\lambda_p}.
 \label{eq:range-scale}
\end{equation}
Setting $H=0$ in
Eq.~\eqref{eq:comb-reach} gives the transmission depth
\begin{equation}
 \Dopt(L)=\Theta_{p,b,\eps}(\sqrt{\ln L})
 \label{eq:time}
\end{equation}
at fixed accuracy and readout error. The constructions
below exhibit the transition to coherent transport.

For a single qubit $A$ and all outputs $B$ at distance at least $L$
on either or both sides, we have
\begin{equation}
 L\ge4\xi\ln(2\xi)
 \quad\Longrightarrow\quad \rho_{AB|\boldsymbol y}\text{ is separable for every }\boldsymbol y.
 \label{eq:strict-main}
\end{equation}
To see this, replace all reports within distance $2d$ of $A$.
The resulting product contribution dominates $a\omega_{\boldsymbol y}$, where
$a=e^{-R(4d+1)\lambda_p}$, and its $A$ factor is fixed under
conditioning outside this region. On each populated side choose
$K_\star=\lfloor(L-2d-1)/s\rfloor$ disjoint windows outside it.
A fully replaced window on each side isolates $A$ from all of $B$.
A union bound in CP order bounds the remaining relative weight by
$\Pi_\star=2(1-e^{-sR\lambda_p})^{K_\star}$, including with the
anchor fixed. The separable-neighborhood proof in the End Matter
therefore applies whenever $\Pi_\star\le a/3$.
It suffices that $K_\star\ge e^{sR\lambda_p}\ln(6/a)$.
Writing $x=sR\lambda_p$ and rounding upward gives a sufficient
integer distance no larger than
$\xi(2e^{-x}+\ln6+2x)\le4\xi\ln(2\xi)$.
After tracing out the intervening registers, local filters and record
selection preserve separability.

For $p=0.4$, $d=1$ and $R=2$, Eq.~\eqref{eq:strict-main} guarantees separability in every record for $L\ge2183$. Figure~\ref{fig:finite}(a) gives exact thresholds for a specified circuit.

\begin{table}[tb]
\caption{\textbf{Optimal transmission depth on a chain.} Entries denote $\Theta(\cdot)$ scalings at fixed width and Bell fidelity strictly between $1/2$ and one. The constructions use ideal quantum operations and fixed readout error $0<p<1/2$. The same noisy orders are necessary even to retain arbitrarily weak transmitted entanglement, at any positive postselection yield.}
\label{tab:times}
\centering
\renewcommand{\arraystretch}{1.15}
\begin{tabular}{@{}llc@{}}
\toprule
Readout & Protocol & $\Dopt$\\
\midrule
None & Unitary transport & $L$\\
Perfect & Teleportation & $1$\\
\addlinespace[3pt]
Noisy & One terminal round & $\ln L$\\
      & Fixed interaction depth & $\ln L$\\
      & Copies and rereads & $\bm{\sqrt{\ln L}}$\\
\bottomrule
\end{tabular}
\end{table}
The separating windows avoid the outputs throughout. Exact separability
also requires them to avoid the anchor.
Table~\ref{tab:times} summarizes the depth laws. The End Matter proves
Eq.~\eqref{eq:uniform-herald}, with
$s=2d+2$, $K=\lfloor(L-1)/s\rfloor$, and
$\Pi=(1-e^{-\lambda_p Rs})^K$.
For $K\ge1$, accepted fidelity at least $1-\eps$ requires
\begin{equation}
 Rs\ln\frac{1-p}{p}
 \ge\ln K-\ln[-\ln(1-2\eps)].
 \label{eq:Snecessary}
\end{equation}
For $0<\eps\le\eps_0<1/2$, the last nested logarithm involves
$2\eps\le-\ln(1-2\eps)\le C_{\eps_0}\eps$,
where $C_{\eps_0}=-\ln(1-2\eps_0)/\eps_0$.
For $L\ge16$ and $1\le d\le L/16$,
$K\ge(L-1)/(4d)-1\ge L/(8d)$.
Consequently,
\begin{equation}
 R(2d+2)\lambda_p
 \ge\ln\frac{L}{d\eps}-\ln(8C_{\eps_0}).
 \label{eq:tradeoff}
\end{equation}
The additive constant is independent of $p$.
If $d=0$, local operations and classical communication cannot transmit
entanglement. If $R=0$ and $d<L$, the output cannot depend on the input.

For total depth $D$, choose windows of length $2D+2$ and pad to $bD$
slots per site. For $1\le D\le L/16$,
\begin{equation}
 bD(2D+2)\ln\frac{1-p}{p}+\ln D
 \ge\ln\frac L\eps-O_{\eps_0}(1).
 \label{eq:Sduration}
\end{equation}
At fixed $p,b$ its left side is $O(D^2)$. If $D>L/16$ the linear
alternative already holds. Small $L$ is absorbed into the constants.
Thus
$D=\Omega_{p,b,\eps_0}(\min\{L,\sqrt{\ln[(L+1)/\eps]}\})$,
independently of the acceptance probability. A deterministic protocol with
half-diamond error at most $\eps$ also has Bell fidelity at least
$1-\eps$, so the converse applies to that stronger error criterion.
The construction in Sec.~\ref{sm:wire-upper} supplies the matching
upper bounds.

With one terminal readout round, a fixed local circuit is followed by
simultaneous measurement and a product correction channel. Bounded width
gives $R=O(1)$, so Eq.~\eqref{eq:Snecessary} yields
$D=\Omega_p(\min\{L,\ln[(L+1)/\eps]\})$.
For fixed $d\ge1$, $L\ge16d$, and $0<\eps\le\eps_0<1/2$,
Eq.~\eqref{eq:tradeoff} gives
$R=\Omega_{p,d,\eps_0}(\ln[(L+1)/\eps])$.
The repeated-readout ladder below attains the latter bound in
$D=O_p(\ln[(L+1)/\eps])$.

\paragraph*{Unconditional channel and exact transmission.}
Before conditioning on success, each fully replaced window has
probability $(2p)^{Rs}$. Set
\begin{equation}
 \Pi_{\rm av}=\bigl[1-(2p)^{Rs}\bigr]^K,\qquad
 \channel=(1-\Pi_{\rm av})\channel_{\EB}+\Pi_{\rm av}\channel_* .
 \label{eq:unconditional}
\end{equation}
At least one uninformative window occurs with input-independent
probability $1-\Pi_{\rm av}$. Conditioned on this event, the channel is
entanglement breaking by cut deletion. The same statement holds with the complete
record at the receiver, since the Choi operator of each fixed record is
separable. Consequently
$\Fe\le(1+\Pi_{\rm av})/2$ and
$d_\diamond(\channel,\channel_{\EB})\le \Pi_{\rm av}$.
For the binary detector, $\Pi_{\rm av}\le\Pi$.
The unconditional bound uses $\ln[1/(2p)]$ in place of
$\ln[(1-p)/p]$ in Eq.~\eqref{eq:Snecessary}.

Exact accepted transmission requires $d\ge L$ for any $p>0$ and a finite
number of reports. For a fixed record $\boldsymbol y$ containing $n_{\boldsymbol y}$ reports,
let $\mathcal T_{\boldsymbol y}$ be the trace-preserving frozen-controller circuit with
all measurements made nonselective. Positivity of the instrument maps gives
$p^{n_{\boldsymbol y}}\mathcal T_{\boldsymbol y}\preceq_{\rm CP}\channel_{\boldsymbol y}
\preceq_{\rm CP}(1-p)^{n_{\boldsymbol y}}\mathcal T_{\boldsymbol y}$.
For $d<L$ the output of $\mathcal T_{\boldsymbol y}$ is independent of the input,
and its Choi state is product.
Any receiver filter accepting the record $\boldsymbol y$ with positive probability
therefore accepts a positive product contribution from $\mathcal T_{\boldsymbol y}$.
The accepted Choi state cannot be a pure Bell state.

\paragraph*{Uniform readout dependence at fixed accuracy.}
Fix $0<\eps<1/2$. For $\alpha>0$, let
$x_\alpha$ be the unique positive solution of
\begin{equation}
 \alpha x_\alpha^2=\ln(L/x_\alpha).
 \label{eq:information-scale}
\end{equation}
For a fixed strip supporting the construction,
\begin{equation}
 D_{\eps,\bullet}^{\star}(L,p)
 =\Theta_{b,\eps}\!\left(\min\{L,\max\{1,x_\alpha\}\}\right),
 \quad
 \alpha=
 \begin{cases}
 \kappa_p,&\bullet=\mathrm{det},\\
 \lambda_p,&\bullet=\mathrm{her}.
 \end{cases}
 \label{eq:uniform-information-time}
\end{equation}
Here $\mathrm{det}$ denotes transmission with every run kept, and $\mathrm{her}$
denotes heralded transmission at any positive yield. We test the Bell fidelity
of the reference and output, conditioning on acceptance in the heralded case.
The constants are independent of $p,L$. The deterministic construction also
controls half-diamond error.
For $K\ge1$, the exact product bound $n\lambda_p\ge-\ln(2\eps)>0$ absorbs the additive constant in Eq.~\eqref{eq:tradeoff}.

For the deterministic lower bound, set $\delta=1-2\eps$.
When $1\le D\le L/16$, $n=R(2d+2)\le4bD^2$ and
$K\ge L/(8D)$ give
$8b\kappa_pD^2\ge\ln[1+\delta^2L/(8D)]$.
If $x_{\kappa_p}\ge1$ and
$D\le c x_{\kappa_p}$, choose
$c\le\min\{1/32,\delta^2/8,1/(4\sqrt b)\}$.
The right side is then at least $\ln(L/x_{\kappa_p})$,
while the left side is at most half that value, a contradiction.
For $x_{\kappa_p}<1$ one nontrivial layer supplies the lower bound.
For heralding the same comparison uses Eq.~\eqref{eq:Snecessary} and a smaller $c=c_{b,\eps}$.

For achievability take $\ell=t$ to be the next odd integer at least
$C_\eps\max\{1,x_\alpha\}$, with
$C_\eps\ge3$ and
$C_\eps^2\ge1+\ln(2/\eps)/\ln2$.
If $\ell\le L/2$, then
$\alpha\ell^2\ge\ln[2L/(\ell\eps)]$, and the protected wire
has error at most $\eps$. Use majority decoding for
$\alpha=\kappa_p$ and unanimous-report selection for
$\alpha=\lambda_p$. If $\ell>L/2$, exact SWAP transmission takes
$L<2\ell$ layers and preserves the same order bound.
Equation~\eqref{eq:unanimity-uniform} extends the heralded result uniformly to all $0<\eps\le\eps_0<1/2$.
Ideal teleportation has constant depth at $p=0$. At $p=1/2$ the
reports contain no information and linear interaction depth is necessary.

\paragraph*{Separated output regions.}
Let $A$ and $B$ be fixed output regions on opposite sides of an interval
of $r-1$ sites. Start from a state product across sites and discard all
quantum registers outside these regions after the costed protocol.
For every reported record, and after any positive-probability classical
selection and local filters on $A$ and $B$,
\begin{equation}
 \inf_{\sigma\in\operatorname{Sep}(A:B)}
 \frac12\|\rho_{AB}^{\rm acc}-\sigma\|_1
 \le
 \left[1-e^{-\lambda_p R(2d+2)}\right]^{{\lfloor
 (r-1)/(2d+2)\rfloor}}.
 \label{eq:separated-regions}
\end{equation}
Place the windows inside the gap and apply Eq.~\eqref{eq:uniform-herald}
to their cut decomposition. A classical copy of the record can be
retained on either or both sides. Quantum operations on the intervening
registers are included in the protocol depth.

\paragraph*{A bounded region and all distant outputs.}
The strict theorem extends to any fixed output region. Let $A$ be
contained in an interval of $r$ sites,
and let $B$ lie outside the $L$-neighborhood of that enclosing interval,
on either or both sides.
Replace all reports within distance $2d$ of the enclosing interval of
$A$. The product contribution dominates
$a_A\omega_{\boldsymbol y}$, with $a_A=e^{-\lambda_p R(r+4d)}$, and its $A$ factor
remains fixed under outside-flag conditioning. A replaced window on each populated side isolates $A$, leaving
residual relative weight at most $\pi=2(1-e^{-\lambda_p Rs})^{\lfloor(L-2d-1)/s\rfloor}$.
Using Lemma~\ref{lem:one-sided-separable} on the support of the product contribution, the proof of Theorem~\ref{thm:range} applies whenever
$\pi\le a_A/(\dim A+1)$. Thus a sufficient distance is
\begin{equation}
 L\ge 2d+1+s\left\lceil e^{sR\lambda_p}
 \left[\ln(2(\dim A+1))+R(r+4d)\lambda_p\right]\right\rceil.
 \label{eq:bounded-region-range}
\end{equation}
For fixed $r$ and $\dim A$, this is $O_{r,\dim A}(\xi\ln(2\xi))$.
The bound applies to the reduced state on $AB$ after the protocol,
with the intervening registers traced out.

\paragraph*{Reporting capacity and width.}
With ideal readout and all-to-all two-qubit gates, parity feedforward and
exponentially many ancillas suffice to synthesize any unitary exactly in
constant quantum depth~\cite{CaoUnitary2026}. On a strip with $w$ qubits
per site, $t$ readout layers give $R\le wt$ reports per site when each
qubit supplies at most one report per layer.
Equation~\eqref{eq:tradeoff} implies
\begin{equation}
 wt(2d+2)\lambda_p
 \ge\ln\frac{L}{d\eps}-O_{\eps_0}(1),
 \qquad 1\le d\le L/16.
 \label{eq:width-budget}
\end{equation}
In particular, fixed $d,t$ require $w=\Omega_p(\ln L)$ at fixed
accuracy. A transverse copying chain
on a rectangular nearest-neighbor strip takes depth proportional to its
height, so this substitution alone does not give a matching
constant-depth construction as width grows. For a full rectangular
nearest-neighbor strip, Eq.~\eqref{eq:narrow-strip-time} gives matching
bounds when $w=O((\ln L)^{1/3})$. For circuit noise in a fault-tolerant quantum bus, Choe and
K\"onig give an $O(\ln^2 L)$ construction and an $\Omega(\ln L)$ lower
bound on qubits per station~\cite{Choe2024}. On a planar strip of polylogarithmic width, Corollary~8.2.2 of
Ref.~\cite{Harley2026} gives a constant-depth resource whose bulk is
measured in $Z$, followed by a Pauli correction on a retained qubit.
With ideal gates and storage, read each measured bulk qubit an odd
number of times and take the majority. The decoded flips have
independent probability $q_t(p)$. The joint state of the decoded record and
the surviving qubits is the same as for independent $X$ faults on
the measured bulk qubits immediately before ideal $Z$ measurement.
For fixed $p<1/2$ and fixed target Bell fidelity, a constant number of
rounds puts $q_t(p)$ below the threshold of the corollary. The conversion
in Sec.~\ref{sm:operations} gives a complete parity measurement with
constant additional depth at fixed accuracy. Whether logarithmic width suffices at constant depth remains open.

\paragraph*{Different gate and measurement times.}
Suppose a gate layer takes time $\tau_g>0$ and a readout round takes time $\tau_m>0$, with fixed reporting error $p$ per round. For a fixed-width strip supporting the construction, the optimum physical duration satisfies
\begin{equation}
 \Topt(L)=
 \Theta_{p,b,\eps_0}\!\left(
 \min\left\{\tau_gL,\,
 \tau_g+\tau_m+\sqrt{\tau_g\tau_m\ln\frac{L+1}{\eps}}\right\}\right).
 \label{eq:weighted-time}
\end{equation}
For the upper bound, choose odd integers of the orders
\begin{equation}
 \begin{aligned}
 \ell&=\max\left\{3,\sqrt{\frac{\tau_m}{\kappa_p\tau_g}
                         \ln\frac{L+1}{\eps}}\right\},\\
 t&=\max\left\{1,\sqrt{\frac{\tau_g}{\kappa_p\tau_m}
                         \ln\frac{L+1}{\eps}}\right\},
 \end{aligned}
\end{equation}
rounded upward with a sufficiently large constant multiplier. Then $\ell t$ supplies the required logarithmic redundancy, while the cost is $O(\tau_g\ell+\tau_mt)$. Taking the faster of this schedule and coherent SWAP transmission gives the minimum in the bound. If the construction requires $\ell$ comparable to $L$, its predicted cost is already of order $\tau_gL$, so the SWAP fallback preserves Eq.~\eqref{eq:weighted-time}.

For the lower bound, $d\le T/\tau_g$ and $R\le bT/\tau_m$. If the interaction depth is a fixed fraction of $L$, or if no reports are used, coherent causality gives the first alternative. Otherwise Eq.~\eqref{eq:tradeoff} applies. A successful protocol in this regime uses both an interaction and a report, so $T\ge\max\{\tau_g,\tau_m\}$. In particular,
$\ln d\le d\le T/\tau_g\le T^2/(\tau_g\tau_m)$.
Combining $T^2/(\tau_g\tau_m)=\Omega_{p,b,\eps_0}(\ln[(L+1)/\eps])$ with the minimum layer times proves the second alternative. The bound assumes a fixed error per readout round. Changing a detector's integration time may also change $p$ and is a separate optimization.

For the two finite schedules used in Fig.~\ref{fig:reach}(b), the first
depths reaching $5\%$ error take $9\tau_g+3\tau_m$ (distributed) and
$8\tau_g+5\tau_m$ (concentrated). At 13 layers their respective
schedule families achieve $0.20\%$ and $1.63\%$ error. Counting each
SWAP as three CNOTs instead of one gate changes the explicit 12-layer
schedule to 14 layers, compared with the same coherent lower bound
of 4200 layers.

\paragraph*{Unequal operation times on a comb.}
The height-dependent law also extends to physical time. Assume
nondemolition readout is available, with layer costs $\tau_g,\tau_m>0$
and fixed reporting error $0<p<1/2$. Fix $0<\eps<1/2$ and the local
reporting capacity $b$, and write $S=\ln(L+1)$. Then
\begin{equation}
 \Topt(L,H)=
 \Theta_{p,\eps,b}\!\left(
 \min\left\{\tau_gL,\,
 \tau_g+\tau_m+
 \max\left\{(\tau_g^2\tau_m S)^{1/3},
              \sqrt{\frac{\tau_g\tau_m S}{H+1}}\right\}\right\}
 \right).
 \label{eq:weighted-comb-time}
\end{equation}
The constants are independent of $H,\tau_g,\tau_m$. The lower
bound applies to non-entanglement-breaking transmission in a single record, and a
deterministic protocol attains the stated accuracy.

For the lower bound, a protocol of duration $T$ has
$d\le T/\tau_g$ and $R\le bT/\tau_m$.
If $T\ge\tau_gL/32$, the first alternative already supplies the bound.
Otherwise successful transmission requires an interaction and a
report, so $d,R\ge1$ and $T\ge\max\{\tau_g,\tau_m\}$.
The packing condition in Eq.~\eqref{eq:comb-exact-separability}
then holds. Its sufficient separability condition must fail for an
entangled output. Using
$u_d(H),v_d(H)=O[d\min\{d,H+1\}]$ gives
\begin{equation}
 S\le C_p R d\min\{d,H+1\}.
\end{equation}
The logarithms of the packing prefactor and anchor weight are
absorbed because $R d\min\{d,H+1\}\ge d\ge1$.
Substituting the time budgets proves both
$T^3\gtrsim_{p,b}\tau_g^2\tau_m S$ and
$T^2\gtrsim_{p,b}\tau_g\tau_m S/(H+1)$.
Together with the minimum layer times, these give the lower bound.
Bounded $L$ changes only constants.

For achievability, set
\begin{equation}
 x=\max\left\{1,\left(\frac{\tau_m S}{\tau_g}\right)^{1/3},
                  \sqrt{\frac{\tau_m S}{\tau_g(H+1)}}\right\}.
\end{equation}
Choose $\ell=\lceil Cx\rceil$ and take $t$ to be the next odd integer
above $C\max\{1,S/[x\min\{x,H+1\}]\}$.
A sufficiently large $C=C_{p,\eps}$ makes the existing comb
construction accurate enough. Since
$x^2\min\{x,H+1\}\ge(\tau_m/\tau_g)S$,
its cost is $O_{p,\eps}(\tau_gx+\tau_m)$, including rounding
and protected input injection. If the intervals do not fit,
$L<8\ell$, SWAP transmission costs $O(\tau_gL)=O_{p,\eps}(\tau_gx)$.
Otherwise choose the faster of the protected construction and SWAP
transmission. This proves Eq.~\eqref{eq:weighted-comb-time}.

When neither minimum layer costs nor coherent transport dominate,
the long-arm regime begins at
$H+1\gtrsim_{p,\eps}[(\tau_m/\tau_g)\ln(L+1)]^{1/3}$.
For fixed layer times, the corresponding large-distance time scales
are $(\tau_g\tau_m\ln L)^{1/2}$ on the chain and
$(\tau_g^2\tau_m\ln L)^{1/3}$ on the long-arm comb. Fresh-probe readout gives a corresponding upper bound
when its copying and reset costs are included in the time per round.

\paragraph*{Physical duration of the parity operation.}
At fixed $p$, $\eps<1/2$, and local capacity, the nondestructive parity
measurement has optimum
\begin{equation}
 \begin{split}
 T_{\eps,ZZ}^{\star}(L,H)=\Theta_{p,\eps,b}\!\Bigg(
 \tau_m+\min\Bigg\{\tau_gL,\ \tau_g+
 \max\bigg\{&[\tau_g^2\tau_m\ln(L+1)]^{1/3},\\[-2pt]
 &\sqrt{\frac{\tau_g\tau_m\ln(L+1)}{H+1}}\bigg\}\Bigg\}\Bigg).
 \end{split}
 \label{eq:weighted-parity-time}
\end{equation}
The constants are uniform in positive $\tau_g,\tau_m$ and integer $H\ge0$.
The returned bit is an actual output of the classical controller,
computed from its reports and independent randomness. If there are no
reports, its distribution is independent of the data input. Comparing
one even-parity and one odd-parity $Z$ eigenstate then gives worst-case
half-diamond error at least $1/2$. Hence a successful implementation
must have duration at least $\tau_m$.

The parity-to-Bell test in Sec.~\ref{sm:operations} also gives the
Bell-generation lower bound used in Eq.~\eqref{eq:weighted-comb-time}.
The final record-dependent correction is local and introduces neither
a new report nor spreading of quantum support, so the same resource
count applies to the test. Combining that bound with $T\ge\tau_m$
proves the lower bound in Eq.~\eqref{eq:weighted-parity-time}.

For the upper bound, the protected Bell construction and the two
protected endpoint readouts cost
$O_{p,\eps,b}(\tau_g+\tau_m)$ plus the maximum term in
Eq.~\eqref{eq:weighted-parity-time}. Alternatively, a traveling ancilla
accumulates parity by a CNOT from the first data qubit, SWAPs along the
backbone, and a CNOT from the second data qubit. Reading this pointer
with $O_{p,\eps}(1)$ repeated reports implements the parity instrument
in $O_{p,\eps}(\tau_g L+\tau_m)$ time. Taking the faster construction
proves the stated optimum. The final readout cost distinguishes this
operation from measurement-free coherent transmission.

\FloatBarrier
\section{Quantum transmission with spatial copies and repeated readout}
\label{sm:wire-upper}

\paragraph*{Block layout and Bell links.}
Choose odd integers $\ell\geq3$ and $t\geq1$. If $L<2\ell$, use
successive SWAP gates along the main row to transmit the input exactly
in depth $O(L)$, with no measurements. Otherwise let
$M=\lfloor L/\ell\rfloor$. Partition sites $1,\ldots,L$ into $M$
consecutive blocks of length $\ell$, adding the remaining
$L-M\ell$ sites to the last block. Every block therefore has room
for $\ell$ spatial copies. Denote the left and right main-row endpoints
of block $j$ by $A_j$ and $B_j$.

In parallel in all blocks, prepare
$\lvert\Phi^+\rangle_{A_jB_j}$ by creating a Bell pair on the first
two sites and moving its second half to $B_j$ by adjacent SWAPs.
This takes $O(\ell)$ nearest-neighbor layers, including a constant
factor if each SWAP is decomposed into CNOTs. All block interiors
return to $\lvert0\rangle$. No copy or measurement is used to create
these links.

Let $A_{\rm in}$ be the input at site $0$. Apply the Bell-basis transformations
\begin{equation}
 \mathrm{CNOT}_{A_{\rm in}\to A_1},\ H_{A_{\rm in}};
 \qquad
 \mathrm{CNOT}_{B_j\to A_{j+1}},\ H_{B_j}
 \quad(1\leq j<M).
 \label{eq:wire-bell-transform}
\end{equation}
These disjoint transformations measure $2M$ Bell bits, including the
two injection bits. The output $B_M$ is kept.

\paragraph*{Explicit copying paths on four rows.}
Each site is a column of four qubits in rows $0,1,2,3$.
The links and input occupy row $0$.
Only rows $1,2$ are needed away from the first block. Protect each
Bell-measurement qubit with exactly $\ell$ distinct ancillary copies, as follows.
\begin{enumerate}
\item Copy $A_{\rm in}$ into row $3$ at site $0$ through the two initially
zero intermediate qubits at that site. Label the two intermediate qubits $a,b$ and the target $c$, in
rows $1,2,3$, respectively. Apply
\begin{equation}
 \mathrm{CNOT}_{A_{\rm in}\to a},\quad
 \mathrm{CNOT}_{a\to b},\quad
 \mathrm{CNOT}_{b\to c},\quad
 \mathrm{CNOT}_{a\to b},\quad
 \mathrm{CNOT}_{A_{\rm in}\to a}.
\end{equation}
This leaves only $c$ carrying the copied computational-basis value
and restores $a,b$ to zero. Propagate $c$ rightward along row $3$
through sites $0,\ldots,\ell-1$.
\item For each measured right endpoint $B_j$, $j<M$, first copy
vertically to row $2$. Using the initially zero intermediate row-$1$
qubit $a$ and row-$2$ qubit $b$ at the same site, apply
$\mathrm{CNOT}_{B_j\to a}$,
$\mathrm{CNOT}_{a\to b}$,
$\mathrm{CNOT}_{B_j\to a}$.
Then propagate leftward through the last $\ell$ sites of its own
block along row $2$.
\item Copy each $A_j$, $1\leq j\leq M$, vertically to row $1$
and then rightward through the first $\ell$ sites of its own block.
\end{enumerate}
Perform the vertical operations in this order, then propagate along
all three rows in parallel. The paths are disjoint within each row, so
copying takes $O(\ell)$ layers. Each path implements
$\ket{s}\ket{0^\ell}\mapsto\ket{s}\ket{s^\ell}$ on a qubit to be
measured in $Z$. Measuring the copies implements that projection and
the original pointer can be discarded without readout.

\paragraph*{Stable pointers and fresh probes.}
Destructive detection is also possible with a stable pointer $C$.
Couple it by $\mathrm{CNOT}_{C\to A}$ to an adjacent probe in
$\ket0$, measure the probe in $Z$, and reset it. The pointer instrument
is $\rho\mapsto P_s\rho P_s$ with $P_s=\ket{s}\bra{s}$, including for
an entangled initial pointer. Later rounds read the same value.
All pairs operate in parallel, adding one probe per pointer and one
copying, measurement, and reset layer per round. With ideal pointer
storage, copying, and reset this preserves $D=O(\ell+t)$ and gives
$\ell t$ independent reporting errors per Bell bit.

\paragraph*{Repeated readout and channel error.}
With nondemolition readout, measure each spatial copy $t$ times in $Z$.
Since $P_s^t=P_s$, the extra observations leave the same projection on
the data. The majority of the odd number $\ell t$ of reports errs with
probability
\begin{equation}
 q=q_{\ell t}(p):=
 \Pr\{\operatorname{Bin}(\ell t,p)>\ell t/2\}
 \leq e^{-\kappa_p\ell t},
 \qquad \kappa_p=-\ln(2\sqrt{p(1-p)}).
 \label{eq:wire-majority}
\end{equation}
The decision-error variables are independent across measured qubits and
independent of their true values, including when conditioned on the
entire ideal quantum measurement record.

Label the outcomes of Bell measurement $j=0,\ldots,M-1$ by
$(z_j,x_j)$, with $z_j$ the transformed control outcome and $x_j$
the transformed target outcome in Eq.~\eqref{eq:wire-bell-transform}.
The standard teleportation identity, iterated through the links,
gives the unnormalized operator from $A_{\rm in}$ to $B_M$ for each outcome string as
\begin{equation}
 K_s=2^{-M}e^{i\theta_s}
 X^{\bigoplus_jx_j}Z^{\bigoplus_jz_j}.
 \label{eq:wire-kraus}
\end{equation}
This holds on any reference-entangled input. After correction using the
decoded bits, the remaining Pauli exponents are parities of their
independent majority errors $E_{x,j},E_{z,j}$, giving
\begin{equation}
 \mathcal N(\rho)=
 \mathbb E\!\left[X^{E_X}Z^{E_Z}\rho Z^{E_Z}X^{E_X}\right],
 \qquad
 E_X=\bigoplus_{j=0}^{M-1}E_{x,j},\quad
 E_Z=\bigoplus_{j=0}^{M-1}E_{z,j}.
 \label{eq:wire-pauli-channel}
\end{equation}
The two parities $E_X,E_Z$ are independent, and each equals zero with
probability $[1+(1-2q)^M]/2$. Thus the entanglement fidelity and half-diamond
error are
\begin{align}
 \Fe(\mathcal N)&=
 \left[\frac{1+(1-2q)^M}{2}\right]^2,
 \label{eq:wire-fidelity}\\
 d_\diamond(\mathcal N,\id)
 &=1-\Fe(\mathcal N)\leq2Mq
 \leq\frac{2L}{\ell}e^{-\kappa_p\ell t}.
 \label{eq:wire-exact-error}
\end{align}
A Bell input distinguishes the four Paulis, attaining the triangle
inequality bound on the diamond distance. For $L=10^6$, $\ell=t=3$
give $M=333\,333$ and a final block of four sites. At $p=0.01$,
$q\simeq1.22\times10^{-8}$ and $\Fe\simeq0.9919$.

\paragraph*{Readout estimate for the experiment of Smith et al.}
In the unprotected circuit with 100 intermediate qubits~\cite{Smith2026}, the bit correction uses 50 fusion outcomes and the phase correction uses all 100 subsequent $X$-basis outcomes.
We model each report as independently flipped with its site's calibrated error probability $p_i$, with all other operations and endpoint measurements ideal.
For either set of reports $S$, the probability of a correct parity is $[1+\prod_{i\in S}(1-2p_i)]/2$. The Bell fidelity is the product of these two probabilities.
We matched the site colors along the path in Fig.~13 of Ref.~\cite{Smith2026} to the plotted cumulative distribution of \texttt{measure\_2} errors, the readout used for mid-circuit measurements.
Allowing for either path orientation and for sites indistinguishable on the color scale gives fidelities from $0.505$ to $0.530$.
In every case the corrected $XX$ correlation is $0.32$, while the $ZZ$ correlation lies between $0.53$ and $0.61$, so the phase correction, which collects twice as many reports, is the weaker half.
The errors along the chain range from $0.10\%$ to $6.4\%$. Over the intermediate sites their median is $0.34\%$ and their mean $0.56\%$, since five sites above $2\%$ carry a third of the total. A uniform error equal to the mean gives $0.52$, whereas the median would give $0.64$.

\begin{figure}[t]
\centering
\includegraphics[width=5.0in]{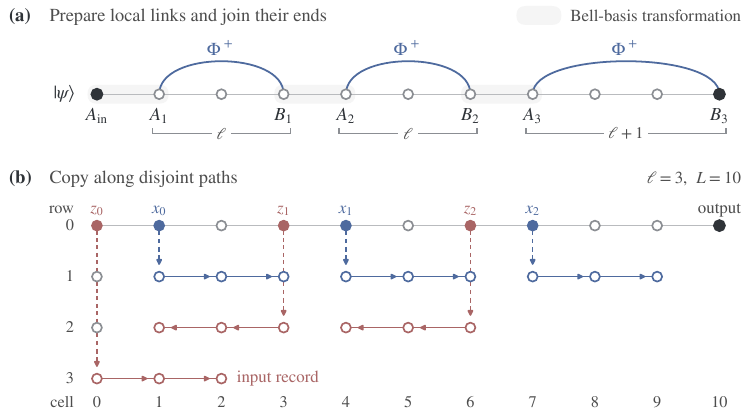}
\caption{\textbf{Four-row transmission with disjoint copying paths.}
(a) Arcs mark local Bell links, shading marks Bell-basis transformations,
and brackets give the block lengths. (b) For $\ell=3$ and $L=10$, the
pointers are copied along disjoint paths. Blue paths carry $x$
values and red paths carry $z$ values, with row 3 reserved for $z_0$.
Dashed vertical transfers precede horizontal copying and restore borrowed
intermediate qubits to zero. Each copy is read $t$ times. Unused ancillary qubits are omitted. The final block has four sites,
leaving enough space for every copy without measuring the output.}
\label{fig:Swire}
\end{figure}

\paragraph*{Resources and matching reach.}
For the layout in Fig.~\ref{fig:Swire}, the interaction depth is
$d=O(\ell)$, the total number of raw reports
at one site is at most $R=3t$, and the total depth is
$D=O(\ell+t)$. Only the injection region needs three copies at one site. Bulk sites
have at most two, so their report budget is $2t$. There are
$2M\ell$ measured ancillas, each read $t$ times, and no other reported
outcomes. Taking $\ell,t$ to be the next odd integers above
\begin{equation}
 \max\!\left\{3,
 \sqrt{\kappa_p^{-1}\ln\frac{2(L+1)}{\eps}}\right\}
\end{equation}
proves the upper bound in Eq.~\eqref{eq:time}, using the SWAP fallback whenever
$L<2\ell$.

For Bell-pair distribution without an input, omit injection and keep
$A_1,B_M$ as outputs. A final nearest-neighbor SWAP moves $A_1$ from site $1$ to the unused qubit at site $0$ when that endpoint is prescribed. This costs constant depth. Only rows $0,1,2$ are needed, and the exact
formula in Eq.~\eqref{eq:wire-fidelity} has $M-1$ in place of $M$.

\paragraph*{Constant interaction depth with repeated readout.}
A two-row nearest-neighbor strip realizes the constant-depth endpoint of
the tradeoff. Label the two qubits at site $i$ by $a_i,b_i$, with their
upper and lower row assignments exchanged on alternating sites. Then
$a_i$ and $b_{i+1}$ are horizontal nearest neighbors, and all pairs
$(a_i,b_{i+1})$, $0\le i<L$, are disjoint. Prepare a Bell pair on each of
these pairs in parallel. Put the unknown input in $b_0$ and apply the
Bell-basis transformation to $(b_i,a_i)$ within each site
$0\le i<L$. Read each of these $2L$ qubits in $Z$ for an odd number $t$
of rounds, majority-decode each bit, and apply the corresponding Pauli
correction to $b_L$. All premeasurement operations have constant depth
using horizontal and vertical edges of the rectangular graph.
If $q_t(p)$ is the majority error for $t$ independent binary reports, the
exact channel is the same Pauli channel as in the block construction,
with $M=L$ and $q=q_t(p)$. Hence
$1-\Fe\le2Lq_t(p)\le2L\exp(-\kappa_p t)$, also bounding the
half-diamond error. The largest number of reports at a site is $2t$ and
the total depth is $t+O(1)$. Choosing
$t=O_p(\ln[(L+1)/\eps])$ therefore attains the logarithmic readout
requirement at constant interaction depth without spatial copying.

\FloatBarrier
\section{Unanimous reports and the noise dependence of heralded transmission}
\label{sm:unanimity}

The transmission circuit has $M$ Bell measurements, including input
injection. To select a run, accept each Bell bit only when all its
$m=\ell t$ reports agree, and accept the run only when every Bell bit
passes. This rule and majority decoding have been tested with repeated
mid-circuit measurements~\cite{Koh2026} and with ancilla-encoded
readout~\cite{Czaban2026}. For either true value of one bit,
\begin{equation}
 a_m=(1-p)^m+p^m,\qquad
 q_m^{\rm un}=\frac{p^m}{a_m}
     =\frac{1}{1+e^{m\lambda_p}},
 \label{eq:unanimity-bit}
\end{equation}
where $a_m$ is its acceptance probability and $q_m^{\rm un}$ its conditional
decoding error. Unlike majority decoding, this rule does not require
$m$ to be odd.

Both $a_m$ and $q_m^{\rm un}$ are independent of the true bit. Conditioned on acceptance, the true Bell outcomes are still jointly
uniform and the residual errors are independent, each with probability $q_m^{\rm un}$. Substitution in
Eqs.~\eqref{eq:wire-pauli-channel}--\eqref{eq:wire-exact-error} gives an
input-independent yield and a normalized Pauli channel,
\begin{equation}
 \begin{aligned}
 a&=a_m^{2M},\\
 \Fe^{\rm acc}
  &=\left[\frac{1+(1-2q_m^{\rm un})^M}{2}\right]^2
    =\left[\frac{1+\tanh(m\lambda_p/2)^M}{2}\right]^2,\\
 d_\diamond(\channel_{\rm acc},\id)
  &=1-\Fe^{\rm acc}
    \le 2M e^{-m\lambda_p}.
 \end{aligned}
 \label{eq:unanimity-channel}
\end{equation}
Thus $\lambda_p\ell t\ge\ln[2L/(\ell\eps)]$ suffices for error $\eps$
on the stated block layout, with $d=O(\ell)$, $R\le3t$, and $D=O(\ell+t)$.

\paragraph*{Average output and the best complete record.}
The wire also distinguishes loss of entanglement in the averaged output
from loss in every reported record. The ideal Bell bits are jointly uniform
and independent of the input, by Eq.~\eqref{eq:wire-kraus}. For a complete
record containing $k$ ones among $m$ reports of one bit, its optimal
posterior error is $[1+\exp(\lambda_p|m-2k|)]^{-1}$. Hence every bit has
posterior error at least
$q_m^{\rm un}=[1+\exp(m\lambda_p)]^{-1}$, with equality for unanimous reports.
At fixed record the posterior errors are independent. For each correction
parity, the bias in favor of the XOR of the individual majority decisions
is the product of their nonnegative posterior biases. This XOR is therefore
an optimal parity decision. The bit and phase parities use disjoint sets
of Bell bits, so the two decisions together maximize the corrected Bell
weight. Its largest possible value occurs for unanimous reports on each
bit, giving the fidelity in
Eq.~\eqref{eq:unanimity-channel}. A two-qubit Bell-diagonal state is
separable if and only if its largest eigenvalue is at most $1/2$.
Consequently, the first number of Bell measurements for which all complete
records of this wire are entanglement breaking is
\begin{equation}
 M_{\rm EB}^{\rm all}
 =\left\lceil
 \frac{\ln(\sqrt{2}-1)}{\ln(1-2q_m^{\rm un})}
 \right\rceil.
 \label{eq:wire-all-record-range}
\end{equation}
The same expression with the majority error $q$ from Eq.~\eqref{eq:wire-majority}
gives the threshold for the majority-corrected channel after averaging
the detailed reports. For one reading per bit, $q=p$ in every record. The threshold in
Eq.~\eqref{eq:wire-all-record-range} is asymptotic to
$-\ln(\sqrt{2}-1)/(2p)$ as $p\to0$ and equals $44$ at $p=0.01$. A pair-distribution chain then uses
$M+1=45$ short Bell pairs. At $p=0.01$ and $m=3$, the averaged wire first
becomes entanglement breaking at $M=1479$. A unanimous record still
leaves an entangled pair up to $M=427597$. From $M=427598$ onward,
every record is separable even after endpoint-local filtering.
At $M=427597$, accepting every string of unanimous triples gives log yield
$2M\ln[(1-p)^3+p^3]\simeq-25784$. One specified string has probability
$\{[(1-p)^3+p^3]/2\}^{2M}$. All these strings yield the same corrected Bell fidelity.
Figure~\ref{fig:finite}(a) converts the largest Bell weight of each
state to concurrence. Counting all records that leave the endpoints entangled
gives an entangled-record probability of $48\%$ at $M=1479$ and $0.11\%$ at $M=2000$. With three reports,
a group splits two votes to one with probability $3p(1-p)$, in which
case its posterior bit error is $p$. For either Pauli correction the
number $i$ of split groups is thus binomial with parameters $M$ and
$3p(1-p)$. Its conditional parity bias is
$v_i=(1-2p)^i(1-2q_3^{\rm un})^{M-i}$, where $q_3^{\rm un}$ is the unanimous error
in Eq.~\eqref{eq:unanimity-bit}. For independent split counts $i,j$
in the two corrections, the largest Bell weight is
$(1+v_i)(1+v_j)/4$. Summing their binomial probabilities wherever
this weight exceeds $1/2$ gives the entangled-record probability.

For this fixed acquisition, the same weights determine the best yield
under classical record selection and Pauli correction at a target Bell fidelity. Write
$F_{ij}=(1+v_i)(1+v_j)/4$. To require fidelity at least $f$ in each
retained record, accept precisely the classes with $F_{ij}\ge f$.
For a target on the accepted average, sort the classes by decreasing
$F_{ij}$ and admit them in that order, randomizing at the boundary if
needed. For any given accepted probability the sorted rule has the
greatest Bell weight, since exchanging equal probability mass from a
lower class for a higher one increases that weight. At $M=1479$,
$p=0.01$, and $f=0.95$, the two yields are $5.42\times10^{-33}$
(each record) and $4.36\times10^{-32}$ (accepted average). The latter
can include some records with fidelity below $f$.

\begin{corollary}[Uniform reporting-error dependence]
On the fixed strip supporting this construction, let
$D_{\eps,\mathrm{her}}^{\star}(L,p)$ be the minimum depth for accepted
Bell fidelity at least $1-\eps$ at any positive yield.
For $0<p<1/2$, $0<\eps\le\eps_0<1/2$, and integer $L\ge1$, let
$x>0$ be the unique solution of
\begin{equation}
 \lambda_p x^2=\ln\frac{L}{\eps x},
 \qquad
 D_{\eps,\mathrm{her}}^{\star}(L,p)
   =\Theta_{b,\eps_0}\!\left(\min\{L,\max\{1,x\}\}\right).
 \label{eq:unanimity-uniform}
\end{equation}
The construction assumes ideal quantum operations. The constants are
independent of $p,L,\eps$.
\end{corollary}
\begin{proof}
For the upper bound, take $\ell=t=2\lceil x\rceil+1$.
Then $\ell\ge\max\{3,2x\}$ and $\ell\le2x+3$. If $\ell\le L/2$,
$\ln[2L/(\ell\eps)]=\lambda_p x^2+\ln(2x/\ell)
\le\lambda_p\ell^2$, proving the sufficient condition and
depth $O(\max\{1,x\})$. Otherwise use exact SWAP transmission
in $L$ steps. Since then $L<2\ell\le4x+6$, this fallback also obeys
the claimed upper bound, including small $L$ and $x<1$.

For the lower bound, a successful shortcut has $D\ge1$, $d\le D$,
and $R\le bD$. Put
$C_0=-\ln(1-2\eps_0)/\eps_0$ and choose
$\delta=\min\{1/32,1/\sqrt{8b},1/(8C_0)\}$.
For $1\le D\le L/16$, the record-by-record converse implies
$4b\lambda_p D^2\ge\ln[L/(D\eps)]-\ln(8C_0)$.
If $1\le x\le L$ and $D\le\delta x$, the right side is at least
$\lambda_p x^2$, while the left side is at most $\lambda_p x^2/2$.
If $x\ge L$, then $\lambda_p L^2\le\ln(1/\eps)$. In this case
$D\le\delta L$ would make the right side at least $\ln(1/\eps)$
and the left side at most half that value. Both cases contradict the
converse. Since $\delta\le1/32$, these assumptions ensure
$D\le L/16$ and, when $D\ge1$, $L\ge32$. The case $x<1$ follows
from $D\ge1$. The lower bound is uniform over $0<\eps\le\eps_0<1/2$.
\end{proof}

The implicit logarithm counts the links remaining when coherent
transport becomes competitive. The deterministic result at fixed
accuracy is Eq.~\eqref{eq:uniform-information-time} with $\kappa_p$.

\FloatBarrier
\begin{figure}[tb]
\centering
\includegraphics[width=4.8in]{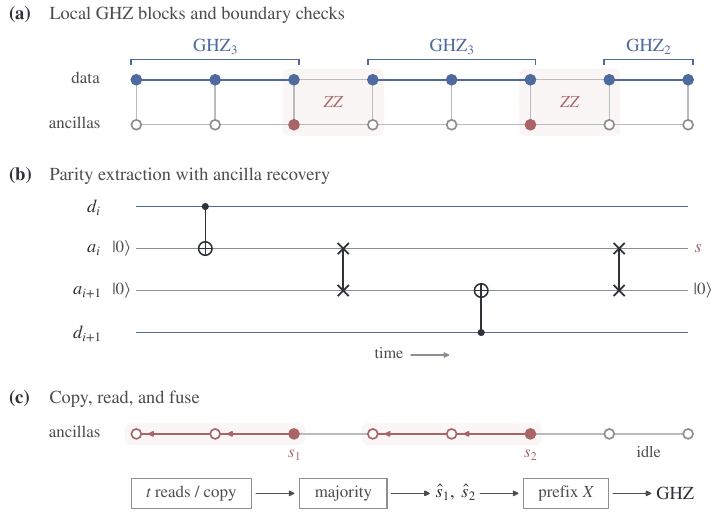}
\caption{\textbf{Fusion of local GHZ blocks.}
(a) Each site holds a data qubit and an ancilla. (b) The four gates
store the boundary parity in $a_i$ and return $a_{i+1}$ to zero.
(c) The ancilla copies in each left block are read $t$ times in parallel.
Majority votes give the boundary parities, whose prefix sums modulo two
determine the blockwise $X$ corrections.}
\label{fig:Sghz}
\end{figure}
\section{GHZ preparation along the backbone}
\label{sm:ghz}
Let $\ket{G_n}=(\ket{0^n}+\ket{1^n})/\sqrt{2}$ be the target on the
$n=L+1$ backbone data qubits. At fixed record $\boldsymbol y$, resolve the
controller seed, replacement flags and fair bits. For the normalized data
state $\rho$ in a contribution with a fully replaced cut,
Eq.~\eqref{eq:adjoint} gives a product marginal on the endpoints. If their local probabilities
of zero are $u,v$, the all-zero and all-one populations obey
$r_0\le uv$ and $r_1\le(1-u)(1-v)$. Positivity then bounds the GHZ coherence,
giving
\begin{equation}
 F_{\rm GHZ}=\bra{G_n}\rho\ket{G_n}\le \tfrac12(\sqrt{r_0}+\sqrt{r_1})^2
 \le \tfrac12\bigl(\sqrt{uv}+\sqrt{(1-u)(1-v)}\bigr)^2\le\tfrac12 .
\end{equation}
The remaining contributions carry at most a fraction $\Pi$ of the record
probability by Eq.~\eqref{eq:branch-domination}. Hence
$F_{\rm GHZ}(\boldsymbol y)\le(1+\Pi)/2$, also after classical selection
of records. Fidelity $1-\eps>1/2$ requires $\Pi\ge1-2\eps$, which with
the cut budgets of Sec.~\ref{sec:comb-proof} gives the depth lower bound
in Eq.~\eqref{eq:comb-depth-lower}.

Koh \emph{et al.}\ fused local GHZ states by parity measurements and
feedforward~\cite{Koh2026}. Waring \emph{et al.}\ used majority votes
over redundant boundary links~\cite{Waring2026}. Here we prepare GHZ
states in parallel on $J=\lceil n/\ell\rceil$ blocks of $\ell\ge2$ sites,
with a possibly shorter last block. Using one ancilla per backbone site,
we store each boundary $ZZ$ parity and copy it through its left block
[Fig.~\ref{fig:Sghz}]. On a comb of height $H$, extend each copy up its
arm by $h=\min\{H,\ell\}$ sites. Reading all copies $t$ times gives
$\ell(h+1)t$ reports per boundary. A majority vote on the largest odd
number $m$ of these reports errs with probability $q\le e^{-\kappa_p m}$.
The prefix parities of the decisions determine blockwise $X$ corrections.
Any incorrect boundary leaves a pattern orthogonal to the target.
Averaging over the independent decoding errors gives
\begin{equation}
 F_{\rm GHZ}=(1-q)^{J-1},\qquad 1-F_{\rm GHZ}\le(J-1)e^{-\kappa_p m}.
\end{equation}
The depth is $O(\ell+h+t)$. Taking $t$ of order $\ell$ gives
$m=\Theta(\ell^2\min\{H+1,\ell\})$, so the choices in
Sec.~\ref{sec:comb-construction} attain the matching depth at fixed accuracy.

\section{Detector likelihoods and rare postselection}
\label{sm:detector}

Suppose a true binary outcome $u$ is reported through a fresh use of a
detector $W(\cdot|u)$, independently of the past given $u$. Write $w_0,w_1$
for the two report densities with respect to the sum $\nu$ of the two laws. Define
\begin{equation}
 \kappa_W=-\ln\int\sqrt{w_0w_1}\,d\nu,\qquad
 \lambda_W=\operatorname*{ess\,sup}_{y}
 \left|\ln\frac{w_0(y)}{w_1(y)}\right|.
 \label{eq:detector-likelihoods}
\end{equation}
If an event has positive probability for just one input, then $\lambda_W=\infty$.
For a finite alphabet the integral reduces to the usual sum of square roots
of probability products. The Bhattacharyya distance $\kappa_W$ measures
distinguishability under repeated readout. The largest log-likelihood
ratio $\lambda_W$ measures how decisive one report can be. It is the
larger of the two directional max-divergences, the Thompson distance
appearing in postselected hypothesis testing~\cite{Regula2024}.
Neither information measure can increase when reports are grouped into bins.
For a bin $B$, Cauchy--Schwarz gives
$\int_B\sqrt{w_0w_1}\,d\nu\le\sqrt{W(B|0)W(B|1)}$.
Summing over bins bounds the Bhattacharyya distance of the digitized detector by $\kappa_W$.
For finite $\lambda_W$, integrating $e^{-\lambda_W}w_1\le w_0\le e^{\lambda_W}w_1$ over each bin gives the same upper bound $\lambda_W$ on its absolute log-likelihood ratio.

For symmetric binary flips the two quantities reduce to $\kappa_p$ and
$\lambda_p$. Related questions concern adaptive use of a noisy measuring
device~\cite{Byrne2026} and communication through a prescribed
postselected channel~\cite{Ji2024}.

The two inputs share a report component of density $\min(w_0,w_1)$.
Its weight $\eta_W=\int\min(w_0,w_1)\,d\nu$ is one minus their total
variation distance. The averaged-channel bound uses $\eta_W$ in place of
the binary replacement probability $2p$.

For an instrument $\{\mathcal M_0,\mathcal M_1\}$, report $y$ induces the
map $\mathcal A_y=\sum_u w_u(y)\mathcal M_u$. Its input-independent
contribution $(\min_u w_u(y))(\mathcal M_0+\mathcal M_1)$ dominates
$e^{-\lambda_W}\mathcal A_y$ in CP order because
$\min_v w_v(y)\ge e^{-\lambda_W}w_u(y)$ for both true values.
Integration gives $\eta_W\ge e^{-\lambda_W}$ for unused slots.
Equations~\eqref{eq:branch-domination} and~\eqref{eq:uniform-herald}
and the product contribution in Theorem~\ref{thm:range} therefore obey
the same bounds with $\lambda_W$ in place of $\lambda_p$ whenever
$\lambda_W<\infty$. For continuous outputs the comparison holds almost
everywhere and after selection of any positive-probability set of records.

Independence cannot be replaced by finite-range correlations and full
support. For one type of Bell report, let the error at join $i$ be
$E_i=B_i\oplus B_{i+1}$, with independent
$B_i\sim\operatorname{Bernoulli}(\theta)$ and $0<\theta<1/2$.
Collections separated by more than one join are independent,
so the stationary noise is one-dependent. Every finite error string has positive
probability, and each report has error $p=2\theta(1-\theta)$.
Nevertheless, its correction-parity error is
$\bigoplus_{i=1}^M E_i=B_1\oplus B_{M+1}$, also of probability $p$,
independently of $M$. Use independent copies of this process for the two
Bell-bit types and for each readout round. Majority decoding of each
\emph{whole correction parity} then reaches any fixed fidelity using a
fixed number of rounds at every distance. Thus finite correlation length
and full support alone do not imply the transmission bound.

Some correlated errors still obey the strict bound. Suppose each run
draws a flip rate $r\in[p_{\min},1/2]$ independently of the input, with
$p_{\min}>0$. Conditional on $r$, the flips are independent. The local
comparison in Eq.~\eqref{eq:local-order} holds for every such $r$ with the
common factor $e^{-\lambda_{p_{\min}}}$. Under the corresponding cut
criterion, each reported record leaves a mixture of separable states.
Observing the record changes the posterior weights of the rates $r$.

\paragraph*{Optimal depth for general detectors.}
Fix the detector, an accuracy $0<\eps<1/2$, and a strip of
fixed width supporting the transmission construction. With ideal quantum
operations, fresh independent reports, and arbitrary positive yield,
Eq.~\eqref{eq:detector-likelihoods} gives
\begin{equation}
 D_{\rm her}^{\star}(L;W)=
 \begin{cases}
  \Theta(L),&\lambda_W=0,\\
  \Theta_W(\sqrt{\ln L}),&0<\lambda_W<\infty,\\
  \Theta(1),&\lambda_W=\infty.
 \end{cases}
 \label{eq:detector-classification}
\end{equation}
For $\lambda_W=0$, the two report laws coincide. Freezing any record leaves
a local trace-preserving circuit. Transmission then requires linear
depth, which a SWAP circuit attains.

For $\lambda_W=\infty$, the likelihood ratio is unbounded in at least
one direction. Exchanging the two true values if needed, every finite
$T$ admits a set $A_T$ of positive $\nu$-measure on which
$w_0(y)\ge e^T w_1(y)$.
The constant-depth wire has $k=2L$ independent uniform Bell bits,
independently of any reference input [Eq.~\eqref{eq:wire-kraus}].
Accept when all reports lie in $A_T$ and decode them as zero. Then
\begin{equation}
 \begin{aligned}
 q_T&=\frac{W(A_T|1)}{W(A_T|0)+W(A_T|1)}\le e^{-T},\\
 a_T&=\left[\frac{W(A_T|0)+W(A_T|1)}2\right]^k>0.
 \end{aligned}
 \label{eq:detector-one-tail}
\end{equation}
The yield is input-independent, and the accepted Pauli channel has error at most $ke^{-T}$. Choosing $T=\ln(k/\eps)$ proves achievability.

For $0<\lambda_W<\infty$, the record-by-record converse gives the lower
bound. For any detector, likelihood-ratio decoding of $m$ independent
reports of a uniform bit gives
\begin{equation}
 q_m\le\tfrac12e^{-m\kappa_W}.
 \label{eq:detector-repetition}
\end{equation}
Since $\kappa_W>0$ whenever the laws differ, spatial copies and repeated
readout attain square-root logarithmic depth, also for asymmetric
detectors. The Bhattacharyya distance equals the Chernoff information
for binary symmetric readout. For general report laws the Chernoff information lies between
$\kappa_W$ and $2\kappa_W$.

\paragraph*{Uniform information cost for a general detector.}
The deterministic comparison extends beyond binary symmetric readout.
For $0<\kappa_W<\infty$, choose an input-independent report density
$g(y)=e^{\kappa_W}\sqrt{w_0(y)w_1(y)}$.
For either true value $u$, the integral of $g^2/w_u$ on the common
support is at most $e^{2\kappa_W}$. On complete histories of
true outcomes and reports, changing the reporting kernel in window $j$
multiplies the positive history operator by
$\mathcal L_j=\prod_{i\in j}g(y_i)/w_{u_i}(y_i)$.
Dummy reports drawn from $W(\cdot|0)$ pad unused slots and integrate
out without any physical measurement.

Order the slots chronologically, breaking ties in a fixed order.
Given the prior history and the current true outcome $u$, the next
multiplier $g(Y)/w_u(Y)$ has mean one and second moment at most
$e^{2\kappa_W}$. Write $\mathcal L_j(t)$ for the partial product over
the first $t$ slots and $N_j(t)$ for the number of those slots in window
$j$. A slot updates at most one of two disjoint windows, so
$\mathcal L_j(t)\mathcal L_k(t)$ is a mean-one martingale for $j\ne k$.
The second-moment bound makes $e^{-2\kappa_W N_j(t)}\mathcal L_j(t)^2$
a supermartingale starting at one. With $N_j(t)\le n$ this gives
$\mathbb E\mathcal L_j^2\le e^{2\kappa_W n}$.
The changed channel is entanglement breaking by the window geometry.
Applying the trace-norm mixture estimate to any reference--input state
and taking the supremum gives
\begin{equation}
 \inf_{\mathcal S\in\mathrm{EB}}
 d_\diamond(\channel,\mathcal S)
 \le\frac12\sqrt{\frac{e^{2\kappa_W R(2d+2)}-1}{K}}.
 \label{eq:general-statistical-eb}
\end{equation}
The constants in this bound are independent of the adaptive quantum
outcome probabilities. Combining it with Eq.~\eqref{eq:detector-repetition}
proves the deterministic part of Eq.~\eqref{eq:uniform-information-time}
with $\alpha=\kappa_W$, uniformly over detector laws at fixed accuracy.
At $\kappa_W=0$ the laws coincide and coherent transport is necessary.
At $\kappa_W=\infty$ their supports are disjoint up to null sets, so
constant-depth teleportation suffices.

At nonnegligible yield, the overlap $\eta_W$ suffices. Write
$K=\lfloor(L-1)/(2d+2)\rfloor$. Any receiver-side trace-nonincreasing CP selection with Bell-input
probability $a$ and accepted fidelity at least $1-\eps$ obeys
\begin{equation}
 a(1-2\eps)\le
 \left(1-\eta_W^{R(2d+2)}\right)^K.
 \label{eq:detector-yield-bound}
\end{equation}
Indeed, selection cannot increase the unnormalized weight of the
non-entanglement-breaking contribution, while the remaining contribution
has Bell overlap at most half its trace. This includes receiver quantum
filtering. For fixed $0<\eta_W<1$ and $a\ge a_0L^{-\gamma}$, with fixed
$a_0>0$ and $\gamma\ge0$, the inequality and $d\le D$, $R\le bD$ force
$D=\Omega_W(\sqrt{\ln L})$. Equation~\eqref{eq:detector-repetition}
provides a deterministic matching construction.

\paragraph*{Mean depth to completion.}
A small success probability also costs depth if rejected stages are
included. Fix a detector with distinct but overlapping report laws,
$0<\eta_W<1$, and a target Bell fidelity $1-\eps>1/2$.
Allow the protocol to retain quantum memory across stages and stop at
a record-dependent depth $T$. Suppose it completes almost surely,
its final averaged Bell fidelity is at least $1-\eps$, and
$\overline T=\mathbb E[T]<\infty$, with both expectation and fidelity evaluated
on the Bell input. All quantum operations and raw reports, including
those in rejected stages, are counted. Then
\begin{equation}
 \inf\mathbb E[T]=\Theta_W(\sqrt{\ln L})
 \label{eq:mean-completion-time}
\end{equation}
for fixed accuracy and width supporting the construction.

To prove the lower bound, set $c=2/(1-2\eps)$ and truncate the protocol
at $\lceil c\overline T\rceil$, accepting only completed histories. Markov's
inequality bounds the omitted probability by $1/c$. If $\omega_\le$
is the resulting unnormalized Bell output, then
\begin{equation}
 2\tr(\Phi^+\omega_\le)-\tr\omega_\le
 \ge 1-2\eps-\frac1c=\frac{1-2\eps}{2}.
 \label{eq:mean-truncation}
\end{equation}
Indeed, removing a set of probability $\delta$ can reduce this linear
Bell advantage by at most $\delta$. For the truncated protocol, the
unconditional decomposition bounds this accepted advantage by the
non-entanglement-breaking weight
$\Pi_{\rm av}=(1-\eta_W^{R(2d+2)})^K$. Selection cannot increase that weight, so
$\Pi_{\rm av}\ge2\tr(\Phi^+\omega_\le)-\tr\omega_\le\ge(1-2\eps)/2$.
Using $d\le\lceil c\overline T\rceil$ and
$R\le b\lceil c\overline T\rceil$ now proves
$\overline T=\Omega_W(\sqrt{\ln L})$. The deterministic construction based on
Eq.~\eqref{eq:detector-repetition} attains the upper bound. The lower bound uses a fixed Bell-fidelity advantage on the Bell
input, so it also applies to worst-case expected depth.

The readout dependence of this mean-depth law is also uniform.
For the truncated protocol, an entanglement-breaking comparison channel
has nonpositive accepted Bell advantage. Equation~\eqref{eq:general-statistical-eb}
therefore implies
$(1-2\eps)/2\le\sqrt{[e^{2\kappa_W n}-1]/K}$.
The lower-bound argument of Eq.~\eqref{eq:uniform-information-time}
applies with $\alpha=\kappa_W$, and the deterministic wire attains it.
For $x_\alpha=O(1)$, the bound $\mathbb E[T]\ge1-2\eps$ from separability
of zero-layer outputs absorbs the truncation ceiling.
Charging failed
stages restores the reliable-transmission information cost even when
arbitrarily rare fixed-depth selection is possible.

Full support alone does not imply the arbitrary-yield, fixed-depth
bound for an unbounded analog record. Consider the idealized detector
\begin{equation}
 Y=(-1)^u\mu+\sigma Z,\qquad Z\sim\mathcal N(0,1),
 \qquad \mu,\sigma>0 .
 \label{eq:analog-detector}
\end{equation}
Here $\kappa_W=\mu^2/(2\sigma^2)$ and $\eta_W=2\Phi(-\mu/\sigma)>0$,
where $\Phi$ is the standard normal cumulative distribution function.
The log-likelihood ratio grows linearly with $Y$, giving $\lambda_W=\infty$. Use the constant-depth
neighboring-link teleportation circuit, whose $k=2L$ Bell bits are
independent and uniform, independently of the input state. Decode each
bit by the sign of its analog report and accept only when all reports
satisfy $|Y|\ge y_{\mathrm c}$. For an accepted report, the posterior error is
$[1+\exp(2\mu|Y|/\sigma^2)]^{-1}\le
\exp(-2\mu y_{\mathrm c}/\sigma^2)$. Thus the accepted transmission has
$1-\Fe^{\rm acc}\le k\exp(-2\mu y_{\mathrm c}/\sigma^2)$. Choosing
$y_{\mathrm c}=(\sigma^2/2\mu)\ln(k/\eps)$ gives error at most $\eps$, with
\begin{equation}
 a=\left[\Prb\{|\mu+\sigma Z|\ge y_{\mathrm c}\}\right]^k>0,
 \qquad
 \ln a=-\Theta_{\mu,\sigma}\!\left(
 k[\ln(k/\eps)]^2\right)
 \label{eq:analog-yield}
\end{equation}
in the large-threshold regime. Since the yield is independent of the
input, the accepted operation defines a transmission channel.

This example assigns no extra cost to retaining arbitrarily large analog
values and relies on their ideal Gaussian tails. Any fixed finite digitizer whose bins have
positive probability for both inputs gives $\lambda_W<\infty$ and restores
the arbitrary-yield bound. Retaining only the sign reduces the detector
to binary symmetric readout.

\FloatBarrier
\section{Complementary-basis tests}
\paragraph*{Parity measurement.}
For an arbitrary instrument the corrected Bell test gives the one-sided
bound in Eq.~\eqref{eq:bell-bridge-main}. For the circuit of
Fig.~\ref{fig:finite}(b) one can also recover the operation error
from two local measurement settings. Prepare $\ket{++}$ and record the
returned parity. Measure both data qubits in $Z$ on one set of runs
and in $X$ on another. For odd reported parity the local correction
reverses the signs assigned to $Z_B$ and $Y_B$ and leaves $X_B$ unchanged.
The corrected $ZZ$ and $XX$ correlations give the biases of the bit and phase corrections, respectively. Their success probabilities multiply because the errors are independent, giving Eq.~\eqref{eq:parity-phase-main}. Joint selection on the two sets of resource reports can correlate
the residual bit and phase errors. In that case their joint success
probability replaces the product. For general joint selection, a third local setting measures $Y_A Y_B$.
The identity $\ket{\Phi^+}\bra{\Phi^+}=(\Id+XX-YY+ZZ)/4$ then gives
the Bell fidelity of the corrected pair from the three correlations. For a Bell-diagonal resource selected before the data
arrive and ideal endpoint reports, this joint Bell weight still gives
$1-\eps_{\rm op}$ by the argument of Sec.~\ref{sm:operations}.

\paragraph*{Transmission.}
Let $\omega_{QB}=(\id_Q\otimes\channel)(\Phi^+_{QA})$ be the normalized Choi state of the unconditioned transmission channel. Define
\begin{equation}
 P_Z=\frac{\Id+Z_QZ_B}{2},\qquad
 P_X=\frac{\Id+X_QX_B}{2}.
\end{equation}
These commuting projectors obey $P_ZP_X=\Phi^+_{QB}$ and
$P_Z+P_X-\Id\le\Phi^+_{QB}$.
Prepare each of $\ket0,\ket1,\ket+,\ket-$ at the input and measure the
output in its matching basis. Let $F_Z,F_X$ be the mean correct-label
probabilities within the two bases. Since these input states are real, the Choi identity gives
$\tr(P_Z\omega)=F_Z$ and $\tr(P_X\omega)=F_X$. Consequently
$\Fe\ge F_Z+F_X-1$~\cite{Hofmann2005}. Equation~\eqref{eq:unconditional}
then bounds the mean score over the two bases,
\begin{equation}
 \bar F:=\frac{F_Z+F_X}{2}\le\frac34+\frac{\Pi_{\rm av}}4 .
 \label{eq:test}
\end{equation}
A channel that measures and prepares in the $Z$ basis has $F_Z=1$,
$F_X=1/2$, and attains the classical bound $\bar F=3/4$. The controller must
receive neither the preparation basis nor the input label.
Input preparation and final detector errors must be calibrated separately
from internal reporting errors.
The same witness applies to accepted transmission with input-independent
acceptance probability, using $\Pi$ in place of $\Pi_{\rm av}$.
For input-dependent acceptance the Choi test requires weighting the
basis scores by their acceptance probabilities.

\end{document}